\magnification=1200
\hoffset=.0cm
\voffset=.0cm
\baselineskip=.55cm plus .55mm minus .55mm

%%%%%%%%%%%%%%%%%%%%%%%%%%%%%%%%%%%%%%%%%%%%%%%%%%%%%%%%%%%%%%%%%%%%%%%%%%%%%%%
%
%       Font loading
%
%	The following makes the AmS fonts available
%	(needs the files amssym.def and amssym.tex)
%
\input amssym.def
\input amssym.tex
%
% 
%%%%%%%%%%%%%%%%%%%%%%%%%%%%%%%%%%%%%%%%%%%%%%%%%%%%%%%%%%%%%%%%%%%%%%%%%%%%%%%
%
%
%  Qui di seguito si effettua la definizione della gerarchia matematica
%  per i fonti greci grassetti e per i fonti sans-serif in modo che 
%  venga compiuta la riduzione di corpo a esponente e a pedice quando
%  siano utilizzati in formule matematiche. 
%
%  Modifica effettuata da GS in data  28 novembre 1996
%
%  (Si utilizzano le famiglie numero 13 e numero 14 rispettivamente
%   sperando che non si vada in conflitto con altre famiglie gia` in uso)
%
%                  NON CAMBIARE SENZA GIUSTIFICATO MOTIVO
%                  E/O  SENZA IL PARERE DELL'AUTORE DELLA 
%                  MODIFICA  PRESENTE.
%
%

%%%%%%%%%%%%%%%%%%%%%%%%%%%%%%%%%%%%%%%%%%%%%

\font\grassettogreco=cmmib10
\font\scriptgrassettogreco=cmmib7
\font\scriptscriptgrassettogreco=cmmib10 at 5 truept
\textfont13=\grassettogreco
\scriptfont13=\scriptgrassettogreco
\scriptscriptfont13=\scriptscriptgrassettogreco

% Definition of sansserif fonts

\font\sansserif=cmss10
\font\scriptsansserif=cmss10 at 7 truept
\font\scriptscriptsansserif=cmss10 at 5 truept
\textfont14=\sansserif
\scriptfont14=\scriptsansserif
\scriptscriptfont14=\scriptscriptsansserif

% Definition of script fonts

\font\capital=rsfs10
\font\scriptcapital=rsfs10 at 7 truept
\font\scriptscriptcapital=rsfs10 at 5 truept
\textfont15=\capital
\scriptfont15=\scriptcapital
\scriptscriptfont15=\scriptscriptcapital
\def\scri{\fam=15}

% Definition of Euler  fonts

\font\euler=eusm10
\font\scripteuler=eusm7
\font\scriptscripteuler=eusm5 
\textfont12=\euler
\scriptfont12=\scripteuler
\scriptscriptfont12=\scriptscripteuler

%
%   FINE DEFINIZIONE DELLE FAMIGLIE MATEMATICHE PER I FONTI
%   cmmib10 (greci grassetti)  e  cmss10  (sans-serif)
%
%
%---- qui di seguito c'e` un campione di come vanno utilizzati ----     
%---- i diversi caratteri ESCLUSIVAMENTE IN MODO MATEMATICO.  ----
%---- Nelle definizioni l'unica cosa che puo` essere cambiata  ----
%---- ad arbitrio dell'utente e` il nome delle macro           ----
%---- (e.g. \Gammabf). Tutto il resto va lasciato come sta.   ----
%
%---- Se si vogliono definire anche i caratteri sansserif va  ----
%---- sostituito ovunque \bgr  con \ssm.                      ----
%

%
%
%        End of font calling
%
%%%%%%%%%%%%%%%%%%%%%%%%%%%%%%%%%%%%%%%%%%%%%%%%%%%%%%%%%%%%%%%%%%%%%%%%%%%%%%
%
%
%                           This is for referencing
%
\def\ref#1{\lbrack#1\rbrack}
%
%
%%%%%%%%%%%%%%%%%%%%%%%%%%%%%%%%%%%%%%%%%%%%%%%%%%%%%%%%%%%%%%%%%%%%%%%%%%%%%%%
%
%
%                           Standard Abbreviations
%
\def\dim{{\rm dim}\hskip 1pt}

\def\imag{{\rm Im}\hskip 1pt}

\def\vol{{\rm vol}\hskip 1pt}
\def\ord{{\rm ord}\hskip 1pt}
\def\ker{{\rm ker}\hskip 1pt}
\def\ran{{\rm ran}\hskip 1pt}

\def\tr{{\rm tr}\hskip 1pt}
\def\id{{\rm id}\hskip 1pt}

\def\SL{{\rm SL}\hskip 1pt}
\def\PSL{{\rm PSL}\hskip 1pt}

\def\Fun{{\rm Fun}\hskip 1pt}

\def\Gau{{\rm Gau}\hskip 1pt}

\def\Tor{{\rm Tor}\hskip 1pt}
\def\Princ{{\rm Princ}\hskip 1pt}
\def\Flat{{\rm Flat}\hskip 1pt}
\def\Conn{{\rm Conn}\hskip 1pt}
\def\Harm{{\rm Harm}\hskip 1pt}
\def\Lie{{\rm Lie}\hskip 1pt}
\def\hst1{\hskip 1pt}

%
%
%%%%%%%%%%%%%%%%%%%%%%%%%%%%%%%%%%%%%%%%%%%%%%%%%%%%%%%%%%%%%%%%%%%%%%%%%%%%%%
%
%
%

\hrule\vskip.5cm
\hbox to 16.5 truecm{October 2002  \hfil DFUB 02--11}
\hbox to 16.5 truecm{Version 2  \hfil hep-th/0210244}
\vskip.5cm\hrule
\vskip.9cm
\centerline{\bf ABELIAN DUALITY AND ABELIAN WILSON LOOPS}   
\vskip.4cm
\centerline{by}
\vskip.4cm
\centerline{\bf Roberto Zucchini}
\centerline{\it Dipartimento di Fisica, Universit\`a degli Studi di Bologna}
\centerline{\it V. Irnerio 46, I-40126 Bologna, Italy}
\centerline{\it and }
\centerline{\it INFN, sezione di Bologna}
\vskip.9cm
\hrule
\vskip.6cm
\centerline{\bf Abstract} 
\vskip.4cm
\par\noindent
We consider a pure $U(1)$ quantum gauge field theory on a 
general Riemannian compact four manifold. 
We compute the partition function with Abelian Wilson loop insertions.
We find its duality covariance properties and derive topological 
selection rules. 
Finally, we show that, to have manifest duality, one must 
assume the existence of twisted topological sectors besides  
the standard untwisted one.

\vfill\eject

\vskip .3cm {\bf 1. Introduction and conclusions}

Electromagnetic Abelian duality is an old subject that has fascinated 
theoretical physicists for a long time as a means to explain the quantization 
of electric charge \ref{1,2,3,4} and the apparent absence of magnetic charge
\ref{5,6,7,8,9}. 
Its study has also provided important clues in the analysis of
analogous dualities in supersymmetric gauge theory \ref{10,11},
supergravity \ref{12,13} and string theory \ref{14,15,16,17}.
It is also considerably interesting for the nontrivial
interplay of quantum field theory, geometry and topology
it shows \ref{18,19,20,21}. 
The aim of this paper is to further explore
these latter aspects of Abelian duality as we briefly outline next.
For an updated review of these matters, see for instance 
refs. \ref{22,23,24}.

Consider a pure $U(1)$ gauge field theory on a general Riemannian 
compact four manifold $M$. The Wick rotated action is 
$$
S(A,\tau)={i\over 2}\int_MF_A\wedge *F_A
+{q^2\theta\over 8\pi^2}\int_MF_A\wedge F_A.
\eqno(1.1)
$$
Here, the charge $q$ and the angle $\theta$ are combined as the real and 
imaginary parts of the complex parameter 
$$
\tau={\theta\over 2\pi}+i{2\pi\over q^2}
\eqno(1.2)
$$
varying in the open upper complex half plane $\Bbb H_+$.  
$A$ is the physical gauge field. 
Its field strength $F_A=dA$ satisfies the quantization condition
$$
{q\over  2\pi}\int_\Sigma F_A\in \Bbb Z,
\eqno(1.3)
$$
for any $2$--cycle $\Sigma$. 

The quantization of the gauge field theory is attained as usual by 
summation over all topological classes of the gauge field and by 
functional integration of the quantum fluctuations of the gauge field 
about the vacuum gauge configuration of each class with the gauge group 
volume factored out. In this way, one can compute in principle the 
partition function possibly with gauge invariant insertions.

It is known that the partition function proper $Z(\tau)$ is a modular 
form of weights $\chi+\eta\over 4$, $\chi-\eta\over 4$ of the subgroup 
$\Gamma_\nu$ of the modular group generated by 
$$
\tau\rightarrow -1/\tau, \quad \tau\rightarrow\tau+\nu,
\eqno(1.4)
$$
where $\chi$ and $\eta$ are respectively the Euler characteristic and 
the signature invariant of $M$ and $\nu=1$ if $M$ is a spin manifold and 
$\nu=2$ else \ref{18}. This property of $Z(\tau)$ is what is usually meant by 
Abelian duality. The natural question arises whether the partition function 
with simple gauge invariant insertions exhibits analogous duality covariance 
properties. Specifically, we shall consider the partition function with 
insertion of the Abelian Wilson loop associated to a $1$--cycle $\Lambda$ 
of $M$:
$$
Z(\Lambda,\tau)
=Z(\tau)\left\langle\exp\left(iq\oint_\Lambda A\right)\right\rangle_\tau.
\eqno(1.5)
$$
In due course, we shall discover the following.

$a$) Due to a peculiar combination of the contributions of the torsion 
classical topological classes and the quantum fluctuations in the field 
theory, the partition function $Z(\Lambda,\tau)$ vanishes unless the 
$1$--cycle $\Lambda$ is a boundary. 

$b$) $Z(\Lambda,\tau)$ is a member of a family of partition 
functions $Z_A(\Lambda,\tau)$ mixing under the transformations (1.4).
$Z_A(\Lambda,\tau)$ is of the general form 
$$
Z_A(\Lambda,\tau)=
\exp\left(-{\pi\sigma(\Lambda)\over \imag \tau}\right)F_A(\Lambda,\tau),
\eqno(1.6)
$$
where $\sigma(\Lambda)$ is the renormalized selfenergy of the classical 
conserved current associated to the $1$--cycle $\Lambda$. 
When the $1$--boundary $\Lambda$ satisfies certain conditions, 
$F_A(\Lambda,\tau)$ is the $A$-th component of a vector modular form 
$F(\Lambda,\tau)$ of weights $\chi+\eta\over 4$, $\chi-\eta\over 4$ for the 
subgroup $\Gamma_\nu$. 

$c$) To have manifest duality, one must assume the existence of twisted 
topological sectors besides the standard untwisted one, one for each 
independent value of the index $A$. $Z_A(\Lambda,\tau)$ is the partition 
function of twisted sector $A$.

In a topologically non trivial manifold $M$, the definition of the integral 
$\oint_\Lambda A$ is not straightforward, as the gauge field $A$ is not a 
globally defined $1$--form. We approach this problem using the theory of the 
Cheeger--Simons differential characters. This produces however a family of 
possible definitions of $\oint_\Lambda A$ parameterized by the choices of 
certain background fields. In spite of this, the result of the calculations 
of $Z(\Lambda,\tau)$ does not depend on the choices made as it should. 

This fact is related to the $\Lambda$ selection rules mentioned above. 
$Z(\Lambda,\tau)$ is non zero when $\Lambda$ is a $1$--boundary.
When this happens, the choices entering in the definition of 
$\oint_\Lambda A$ turn out to be immaterial. The proof of this 
intriguing result involves 
an interesting relationship between flat Cheeger--Simons differential 
characters and Morgan--Sullivan torsion invariants.

The physical significance of the twisted topological sectors remains 
to be explored. It seems to indicate that the non perturbative structure of 
electrodynamics might be far richer than thought so far.  

\vskip .3cm {\it Plan of the paper}

In sect. 2, we introduce the necessary topological 
set up. We use this to properly define the Wilson loop corresponding 
to a given $1$--cycle. In sect. 3, we proceed to the calculation 
of the partition function with a Wilson loop insertion and show that it 
vanishes unless the associated $1$--cycle is a boundary. In sect. 4, 
we study the duality properties of the partition function and show the 
existence of twisted topological sectors. Finally,in the appendix, 
we collect some of the technical details of the calculation of the 
partition function. 

\vskip .3cm {\it Conventions and notation}

For a review of the mathematical formalism, see for instance \ref{25}.
For a clear exposition of its field theoretic applications, see \ref{26}.

In this paper, $M$ denotes a compact connected oriented four manifold.

For a sheaf of Abelian groups $\scri S$ over $M$, 
$H^p(M,{\scri S})$ denotes the $p$--th sheaf cohomology group
of $\scri S$ and $\Tor(M,{\scri S})$ its torsion subgroup.
For an Abelian group $G$, $G$ denotes the associated constant sheaf 
on $M$. For an Abelian Lie group $G$, $\underline{G}$ denotes 
the sheaf of germs of smooth $G$ valued functions on $M$.

$C^s_p(M)$, $Z^s_p(M)$, $B^s_p(M)$ denote the groups of smooth 
singular $p$--chains, cycles and boundaries of $M$, respectively,
and $b$ the boundary operator. $H^s_p(M)$ denotes the $p$--th singular 
homology group and $\Tor^s_p(M)$ its torsion subgroup.
For an Abelian group $G$, $C^p_{sG}(M)$, $Z^p_{sG}(M)$, $B^p_{sG}(M)$ denote 
the groups of smooth singular $p$--cochains, cocycles and coboundaries of $M$ 
with coefficients in $G$, respectively, and $d$ the coboundary operator. 
$H^p_{sG}(M)$ denotes the $p$--th singular cohomology group 
with coefficients in $G$ and $\Tor^p_{sG}(M)$ its torsion subgroup.

$C^p_{dR}(M)$, $Z^p_{dR}(M)$, $B^p_{dR}(M)$ denote the groups of 
general, closed and exact smooth $p$--forms of $M$, respectively,
and $d$ the differential operator. $H^p_{dR}(M)$ denotes $p$--th de Rham 
cohomology space. Further, $Z^p_{dR\Bbb Z}(M)$ denote the subgroup of 
closed smooth $p$--forms of $M$ with integer periods and  
$H^p_{dR\Bbb Z}(M)$ the integer cohomology lattice in $H^p_{dR}(M)$. 
$q$ denotes the natural homomorphism of $H^p(M,\Bbb Z)$ into 
$H_{dR}^p(M)$. $b_p$ denotes the $p$--th Betti number.
When $M$ is equipped with a metric $g$, $\Harm^p(M)$
denotes the space of harmonic $p$--forms of $M$ and $\Harm^p_{\Bbb Z}(M)$
the lattice $\Harm^p(M)\cap Z^p_{dR\Bbb Z}(M)$.
$b_2^\pm$ denotes the dimension of the space of (anti)selfdual harmonic 
$2$--forms.

\vfill\eject

\vskip .3cm {\bf 2. $U(1)$ principal bundles, connections and Cheeger--Simons 
characters}

In this section, we review well known facts about $U(1)$ 
principal bundles, connections and Cheeger Simons differential characters, 
which are relevant in the following. 
See ref. \ref{27} for background material.

\vskip .3cm {\it 2.1 Smooth and flat principal bundles}

The quantization of Maxwell theory involves a summation over the topological 
classes of the gauge field. Mathematically, these classes can be 
identified with the isomorphism classes of smooth $U(1)$ principal bundles,
which we describe below. 

The group of isomorphism classes of smooth $U(1)$ principal bundles on $M$,
$\Princ(M)$, can be identified with the $1$--st cohomology of the sheaf 
${\underline U}(1)$:
$$
\Princ(M)=H^1(M,{\underline U}(1)).
\eqno(2.1.1)
$$

There is a well known alternative more convenient characterization of 
$\Princ(M)$ derived as follows. Consider the short exact sequence of sheaves
$$
\matrix{
& & & {}_i & & {}_e & &\cr
0&\rightarrow &\Bbb Z &\rightarrow &\underline{\Bbb R}& 
\rightarrow &{\underline U}(1)&\rightarrow 0,\cr
}
\eqno(2.1.2)
$$
where $i(n)=n$ for $n\in \Bbb Z$ and $e(x)=\exp(2\pi ix)$ for $x\in \Bbb R$.
The associated long exact sequence of sheaf cohomology contains the segment
$$
\matrix{
& & {}_{e_*}& &{}_c & & {}_{i_*} & & & \cr  
\cdots\rightarrow & H^1(M,\underline{\Bbb R}) &\rightarrow & 
H^1(M,{\underline U}(1))&\rightarrow & H^2(M,\Bbb Z)&
\rightarrow & H^2(M,\underline{\Bbb R})& \rightarrow \cdots.\cr
}
\eqno(2.1.3)
$$
Since $\underline{\Bbb R}$ is a fine sheaf, 
$H^p(M,\underline{\Bbb R})=0$ for all $p\geq 1$. Therefore,
$c$ is an isomorphism $H^1(M,{\underline U}(1))\cong H^2(M,\Bbb Z)$.
It follows that
$$
\matrix{
&{}_c&\cr
\Princ(M)&\cong & H^2(M,\Bbb Z).\cr
}
\eqno(2.1.4)
$$
This isomorphism associates to any smooth $U(1)$ principal  bundle $P$
its Chern class $c_P$.

Flat $U(1)$ principal bundles play an important role in
determining the selection rules of the Abelian Wilson loops, as will be 
shown later. It is therefore necessary to understand their 
place within the family of smooth $U(1)$ principal bundle. 

The group of isomorphism classes of flat $U(1)$ principal bundles on $M$,
$\Flat(M)$, can be identified with the $1$--st cohomology of the constant
sheaf $U(1)$:
$$
\Flat(M)=H^1(M,U(1)).
\eqno(2.1.5)
$$

There is an obvious natural sheaf morphism $U(1)\rightarrow{\underline U}(1)$, 
to which there corresponds a homomorphism  $H^1(M,U(1))\rightarrow 
H^1(M,{\underline U}(1))$ of sheaf cohomology. By (2.1.1), (2.1.5), 
this can be viewed as a homomorphism of $\Flat(M)$ into $\Princ(M)$. 
Its image is the subgroup of smooth isomorphism classes of flat principal 
bundles, $\Princ_0(M)$.

On account of (2.1.4), $\Princ_0(M)$ is isomorphic to 
a subgroup of $H^2(M,\Bbb Z)$, which we shall identify next.
Consider the short exact sequence of sheaves
$$
\matrix{
& & & {}_i & & {}_e & &\cr
0&\rightarrow &\Bbb Z &\rightarrow &\Bbb R& 
\rightarrow &U(1)&\rightarrow 0,\cr
}
\eqno(2.1.6)
$$
where $i$ and $e$ are defined as above. 
The associated long exact sequence of sheaf cohomology contains the segment
$$
\matrix{
& & {}_{e_*}& &{}_c & & {}_{i_*} & & & \cr  
\cdots\rightarrow & H^1(M,\Bbb R) &\rightarrow & 
H^1(M, U(1))&\rightarrow & H^2(M,\Bbb Z)&
\rightarrow & H^2(M,\Bbb R)& \rightarrow \cdots.\cr
}
\eqno(2.1.7)
$$
Recalling that $\Tor^2(M,\Bbb Z)=\ker i_*|H^2(M,\Bbb Z)$,
$c$ induces an isomorphism 
$H^1(M, U(1))/$ $e_*H^1(M,\Bbb R)\cong\Tor^2(M,\Bbb Z)$.
Using the \v Cech realization of sheaf cohomology, 
it is easy to see that $H^1(M, U(1))/e_*H^1(M,\Bbb R)$ is isomorphic 
to the image of $H^1(M,U(1))$ in $H^1(M,{\underline U}(1))$.
Therefore, we conclude that
$$
\matrix{
&{}_c & \cr  
\Princ_0(M)&\cong & \Tor^2(M,\Bbb Z).\cr
}
\eqno(2.1.8)
$$

Combining (2.1.4), (2.1.8), we conclude that there is a commutative diagram 
$$
\matrix{
                     &     {}_c       &                             \cr
    \Princ_0(M)      & \rightarrow & \Tor^2(M,\Bbb Z)               \cr
            ~        &             &                                \cr
{}^\subseteq~\downarrow &             & \downarrow~{}^\subseteq     \cr
               ~     &             &                                \cr
    \Princ(M)        & \rightarrow & H^2(M,\Bbb Z),                 \cr
                     &     {}_c       &                             \cr
}
\eqno(2.1.9)
$$
where the lines are isomorphisms.
This describes in some detail the set of $U(1)$ principal  bundles on $M$.

Before proceeding to the next topic, the following remark is in order.
The Chern class $c_P$ of a principal $U(1)$ bundle $P$
belongs by definition to the cohomology group $H^2(M,\Bbb Z)$. 
Another definition identifies the Chern class of $P$ with $q(c_P)$, 
the natural image of $c_P$ in the integer lattice $H_{dR\Bbb Z}^2(M)$ of 
de Rham cohomology. The advantage of the first definition, adopted in this 
paper, is that it discriminates principal bundles differing by a flat 
bundle. The second, though more popular in the physics literature, does not.

\vskip .3cm {\it 2.2 The gauge group}

The fixing of the gauge symmetry is an essential step of the quantization 
of Maxwell theory. Below, we recall the main structural properties of 
the gauge group.

For $P\in\Princ(M)$, the gauge group of $P$, $\Gau(P)$, can be 
identified with the $0$--th cohomology of the sheaf ${\underline U}(1)$:
$$
\Gau(P)=H^0(M,{\underline U}(1)).
\eqno(2.2.1)
$$
Its elements are often called large gauge transformations 
in the physics literature.

The flat gauge group of $P$, $G(P)$, can similarly be identified with the 
$0$--th cohomology of the constant sheaf $U(1)$:
$$
G(P)=H^0(M,U(1)).
\eqno(2.2.2)
$$
Its elements are commonly called rigid gauge transformations.

The natural sheaf morphism $U(1)\rightarrow{\underline U}(1)$ induces 
a homomorphism  $H^0(M,U(1))\rightarrow H^0(M,{\underline U}(1))$ of sheaf 
cohomology, which is readily seen to be an injection. Thus, $G(P)$ is
isomorphic to a subgroup $\Gau_0(P)$ of $\Gau(P)$.

Note that 
$$
G(P)\cong \Gau_0(P)\cong U(1).
\eqno(2.2.3)
$$

$\Gau(P)$ and $G(P)$ or $\Gau_0(P)$ do not depend on $P$. Therefore, 
to emphasize this fact, we shall occasionally denote these groups by 
$\Gau(M)$ and $G(M)$ or $\Gau_0(M)$, respectively.

For $h\in H^0(M,{\underline U}(1))$, define 
$$
\alpha(h)={1\over 2\pi i} h^{-1} dh.
\eqno(2.2.4)
$$
It is straightforward to show that $\alpha(h)\in Z_{dR\Bbb Z}^1(M)$ and that
the map $\alpha: H^0(M,{\underline U}(1))\rightarrow Z_{dR\Bbb Z}^1(M)$
is a group homomorphism with range $Z_{dR\Bbb Z}^1(M)$ and kernel
$H^0(M,U(1))$. Thus, on account of (2.2.1)--(2.2.3), we have the important 
isomorphism
$$
\matrix{
&{}_\alpha & \cr  
\Gau(M)/\Gau_0(M)&\cong & Z_{dR\Bbb Z}^1(M).\cr
}
\eqno(2.2.5)
$$

The counterimage of $B_{dR}^1(M)$ by $\alpha$ is the subgroup $\Gau_c(M)$
of $\Gau(M)$ of the gauge group elements homotopic to the identity. 
Its elements are called small gauge transformations 
in the physics literature. Obviously, $\Gau_0(M)\subseteq\Gau_c(M)$. 
Thus, 
$$
\matrix{
&{}_\alpha & \cr  
\Gau_c(M)/\Gau_0(M)&\cong & B_{dR}^1(M).\cr
}
\eqno(2.2.6)
$$
The quotient $\Gau(M)/\Gau_c(M)$ is the gauge class group.
By the above,
$$
\Gau(M)/\Gau_c(M)\cong H_{dR\Bbb Z}^1(M).
\eqno(2.2.7)
$$

\vskip .3cm {\it 2.3 Connections}

After rescaling by a suitable factor $q/2\pi$, the photon gauge field 
of Maxwell theory can mathematically be characterized as a connection of 
some $U(1)$ principal bundle. Next, we recall the main properties of  
the set of connections of a $U(1)$ principal bundle. 

For any $P\in \Princ(M)$, the family of connections of $P$,
$\Conn(P)$, is an affine space modeled on $C^1_{dR}(M)$. For $A\in \Conn(P)$,
$$
F_A=dA
\eqno(2.3.1)
$$ 
is the curvature of $A$. As well known, $F_A\in Z_{dR\Bbb Z}^2(M)$
and $q(c_P)=[F_A]_{dR}$ (cfr. eq. (2.1.4)).

If $P,~P'\in \Princ(M)$, $A\in \Conn(P)$, $A'\in \Conn(P')$,
then $A+A'\in \Conn(PP')$.
If $P\in \Princ_0(M)\subseteq \Princ(M)$
is flat, then $0\in \Conn(P)$. So, if 
$P\in \Princ(M)$, $P'\in \Princ_0(M)$,
then $\Conn(PP')=\Conn(P)$. In particular, $\Conn(P')=\Conn(1)=C^1_{dR}(M)$.

For $P\in \Princ(M)$, $\Gau(P)$ acts on $\Conn(P)$ as usual, viz
$$
A^h=A+\alpha (h)
\eqno(2.3.2)
$$
for $A\in \Conn(P)$ and $h\in \Gau(P)$ (cfr. eq. (2.2.4)). 
Note that $\Gau_0(P)$ is precisely the invariance subgroup of $A$.

\vskip .3cm {\it 2.4 Cheeger Simons differential characters}

As is well known, if $A$ is a connection of some principal $U(1)$ bundle $P$, 
the line integral $\oint_\Lambda A$ over some closed path cannot be defined 
in the usual naive sense, since $A$ suffers local gauge ambiguities and, thus, 
is not a globally defined $1$--form. 
Nevertheless, one can try to give a meaning to such a formal expression 
modulo integers using the theory of the Cheeger Simons differential 
characters, whose main features are described below \ref{27,28,29,30}.

A Cheeger Simons differential character is a mathematical object having 
the formal properties characterizing the holonomy map of a principal 
$U(1)$ bundle. It has however a somewhat wider scope, since it is defined
for singular $1$--cycles, which are objects more general than closed paths. 
Roughly speaking, we define the formal integral $\oint_\Lambda A$ 
as the logarithm of a suitably chosen differential character
computed at the appropriate $1$--cycle $\Lambda $.

A Cheeger Simons differential character is a group homomorphism
$\Phi:Z^s_1(M)\rightarrow U(1)$ such that there is a $2$--form 
$F_\Phi\in C^2_{dR}(M)$ for which
$$
\Phi(b S)=\exp\left(2\pi i\int_S F_\Phi\right),
\eqno(2.4.1)
$$
for $S\in C^s_2(M)$ . 
The Cheeger Simons differential characters form naturally 
a group $CS^2(M)$.

From (2.4.1), it is simple to see that, for $\Phi\in CS^2(M)$, $F_\Phi\in 
Z_{dR\Bbb Z}^2(M)$ and that the map $F:CS^2(M)\rightarrow Z_{dR\Bbb Z}^2(M)$, 
$\Phi\mapsto F_\Phi$ is a group homomorphism.

To any $\Phi\in CS^2(M)$, there is associated a class 
$c_\Phi\in H^2(M,\Bbb Z)$ 
such that $q(c_\Phi)=[F_\Phi]_{dR}$ defined as follows. 
Since $U(1)\cong \Bbb R/\Bbb Z$ is a divisible group and $Z^s_1(M)$ is a 
subgroup of the free group $C^s_1(M)$, there exists a real cochain $f\in 
C^1_{s\Bbb R}(M)$ such that $\Phi=\exp\left(2\pi if\big|Z^s_1(M)\right)$. 
Then, by (2.4.1), 
$$
\varsigma(S)=f(b S)-\int_S F_\Phi,\quad S\in C^s_2(M),
\eqno(2.4.2)
$$
defines an integer cochain $\varsigma\in C^2_{s\Bbb Z}(M)$. It is readily 
checked that $\varsigma\in Z^2_{s\Bbb Z}(M)$ is an integer cocycle which, 
viewed as a real cocycle, is cohomologically equivalent to $F_\Phi$. 
The choice of $f$ affects $\varsigma$ at most by an integer coboundary. 
Hence, the class $c_\Phi$ of $\varsigma$ in the $2$--nd integer 
cohomology $H^2_{s\Bbb Z}(M)$ is unambiguously determined by $\Phi$. 
The statement then follows from the isomorphism of integer singular and sheaf 
cohomology. It is simple to see that the map $c:CS^2(M)\rightarrow 
H^2(M,\Bbb Z)$, $\Phi\mapsto c_\Phi$ is a group homomorphism.

To any $v\in C_{dR}^1(M)$, there is associated an element $\chi_v\in CS^2(M)$ 
by
$$
\chi_v(\Lambda)=\exp\left(2\pi i\oint_\Lambda v\right),
\quad \Lambda \in Z^s_1(M).
\eqno(2.4.3)
$$
One has $F_{\chi_v}=dv$ and $c_{\chi_v}=0$. Clearly $\chi_v$ depends only 
on the class of $v$ mod $Z_{dR\Bbb Z}^1(M)$ and the map $\chi:
C_{dR}^1(M)/Z_{dR\Bbb Z}^1(M)\rightarrow CS^2(M)$, $[v]\mapsto \chi_v$
is a group homomorphism. When $a\in Z_{dR}^1(M)\subseteq C_{dR}^1(M)$, 
$\chi_a$ depends only on the cohomology class of $a$ in $H_{dR}^1(M)$ mod
$H_{dR\Bbb Z}^1(M)$ and the map $\chi:
H_{dR}^1(M)/H_{dR\Bbb Z}^1(M)\rightarrow CS^2(M)$, $[a]\mapsto \chi_a$
is again a group homomorphism.

The above properties are encoded in the short exact sequences 
$$
\matrix{& & & {}_\chi & & {}_{(c, F)} & & & \cr
0&\rightarrow & H_{dR}^1(M)/H_{dR\Bbb Z}^1(M) &
\rightarrow & CS^2(M) & \rightarrow &A_{\Bbb Z}^2(M)&
\rightarrow & 0,\cr
}
\eqno(2.4.4)
$$
$$
\matrix{
& & & {}_\chi & & {}_c & & & \cr
0&\rightarrow &C_{dR}^1(M)/Z_{dR\Bbb Z}^1(M)&
\rightarrow &CS^2(M) &\rightarrow & H^2(M,\Bbb Z)&
\rightarrow &0. \cr
}
\eqno(2.4.5)
$$
Here, $A_{\Bbb Z}^2(M)$ is the subset of the Cartesian product
$H^2(M,\Bbb Z)\times Z_{dR\Bbb Z}^2(M)$ formed by the pairs 
$(e,G)$ such that $q(e)=[G]_{dR}$.

Before entering the details of the definition of the formal integral 
$\oint_\Lambda A$, with $P\in\Princ(M)$, $A\in\Conn(P)$ and 
$\Lambda\in Z^s_1(M)$, let us recall the properties which it is required 
to have. First, when $\Lambda$ is a boundary, so that $\Lambda= b S$ for 
some $S\in C^s_2(M)$, one has
$$
\oint_\Lambda A=\int_S F_A,\quad \hbox{mod $\Bbb Z$},
\eqno(2.4.6)
$$
where the integral in the right hand side is computed according to the 
ordinary differential geometric prescription. This is a formal generalization 
of Stokes' theorem. Second, for $v\in C^1_{dR}(M)$, the obvious relation 
$$
\oint_\Lambda(A+v)=\oint_\Lambda A+\oint_\Lambda v,
\quad \hbox{mod $\Bbb Z$},
\eqno(2.4.7)
$$
holds, where the second integral in the right hand side is computed 
according to the ordinary differential geometric prescription. 
This property may be called semilinearity. Third, for $h\in\Gau(P)$,
$$
\oint_\Lambda A^h=\oint_\Lambda A \quad \hbox{mod $\Bbb Z$}.
\eqno(2.4.8)
$$
In this way, gauge invariance is ensured. This property, albeit
important, is not independent from the others. 
Indeed, it follows from (2.4.7), (2.3.2)
and the fact that $\alpha(h)\in Z^1_{dR\Bbb Z}(M)$
and, thus, $\oint_\Lambda \alpha(h)\in \Bbb Z$.

Tentatively, for $\Lambda\in Z^s_1(M)$, we define $\oint_\Lambda A$ 
mod $\Bbb Z$ as follows. We consider a character $\Phi\in CS^2(M)$ such that 
$c_\Phi=c_P$ and $F_\Phi=F_A$. As $q(c_P)=[F_A]_{dR}$, the condition 
$q(c_\Phi)=[F_\Phi]_{dR}$ is fulfilled. Then, we set 
$$
\Phi(\Lambda)=\exp\left(2\pi i\oint_\Lambda A\right).
\eqno(2.4.9)
$$
The definition given is however ambiguous. 
Indeed, by the exact sequence (2.4.4),
the character $\Phi$ with the stated properties is not unique, being defined 
up to a character of the form 
$\chi_a$ with $a\in Z_{dR}^1(M)$ defined modulo $Z_{dR\Bbb Z}^1(M)$.
The definition is also not satisfactory, since, apparently, it yields the 
same result for connections differing by a closed form 
$a\in Z_{dR}^1(M)$. 

To solve these problems, we proceed as follows. With some natural criterion,
we fix a reference connection $A_P\in\Conn(P)$ and a fiducial character
$\Phi_P\in CS^2(M)$ such that $c_{\Phi_P}=c_P$ and $F_{\Phi_P}=F_{A_P}$ and 
declare $\oint_\Lambda A_P$ to be given  mod $\Bbb Z$ by the above procedure:
$$
\Phi_P(\Lambda)=\exp\left(2\pi i\oint_\Lambda A_P\right).
\eqno(2.4.10)
$$
Next, for a generic connection $A\in\Conn(P)$, we define a form 
$v_A\in C^1_{dR}(M)$ depending on $A$ by the relation 
$$
A=A_P+v_A.
\eqno(2.4.11)
$$ 
Then, we set 
$$
\oint_\Lambda A=\oint_\Lambda A_P+\oint_\Lambda v_A
\quad \hbox {mod $\Bbb Z$}.
\eqno(2.4.12)
$$
It is easy to check that this definition of $\oint_\Lambda A$ has the 
required properties (2.4.6)--(2.4.8).

Note that $\oint_\Lambda A$ depends on $P$ via its Chern class $c_P$ and 
not simply via $q(c_P)=[F_A]_{dR}$. It is therefore sensitive to torsion. 
By the isomorphism (2.1.8), the torsion part of $c_P$ reflects the flat 
factors of $P$. Thus, $\oint_\Lambda A$ depends explicitly on these latter.

Needless to say, what we have done here is to provide a family of 
definitions of $\oint_\Lambda A$ parameterized by the choices of $A_P$ 
and $\Phi_P$. In the next subsection, we shall devise a way of 
restricting the amount of arbitrariness involved. 

\vskip .3cm {\it 2.5  Background connection and character assignments}

We consider below the group isomorphism that associates to any $c\in 
H^2(M,\Bbb Z)$ the unique (up to smooth equivalence) $U(1)$ principal bundle 
$P_c$ such that $c_{P_c}=c$. This map is the inverse of the isomorphism 
(2.1.4).

A background connection assignment is a map that associates to any
$c\in H^2(M,\Bbb Z)$ a connection $A_c\in\Conn(P_c)$ in such a way that
$$
A_{c+c'}=A_c+A_{c'},\quad c,~c'\in H^2(M,\Bbb Z),
\eqno(2.5.1)
$$
$$
A_t=0,\quad t\in \Tor^2(M,\Bbb Z).
\eqno(2.5.2)
$$
We set $F_c=F_{A_c}$.

A background character assignment compatible with a background connection 
assignment $c\mapsto A_c$ is a map that associates to any
$c\in H^2(M,\Bbb Z)$ a character $\Phi_c\in CS^2(M)$ such that
$c_{\Phi_c}=c$ and $F_{\Phi_c}=F_c$ and that
$$
\Phi_{c+c'}=\Phi_c\cdot\Phi_{c'},\quad c,~c'\in H^2(M,\Bbb Z).
\eqno(2.5.3)
$$

A background connection assignment $c\mapsto A_c$ and a compatible 
background character assignment $c\mapsto \Phi_c$ can be constructed as 
follows. 
Let $f_r$, $r=1,\ldots, b_2$ and $t_\rho$, $\rho=1,\ldots, t_2$ 
a set of independent generators of $H^2(M,\Bbb Z)$, where the $f_r$ are 
free and the $t_\rho$ are torsion of order $\kappa_\rho$.
Every $c\in H^2(M,\Bbb Z)$ can be written uniquely as
$$
c=\sum_rn^r(c)f_r+\sum_\rho k^\rho(c)t_\rho,
\eqno(2.5.4)
$$
for certain $n^r(c)\in\Bbb Z$ depending linearly on $c$
and $k^\rho(c)=1,\ldots,\kappa_\rho-1$ depending linearly on $c$
modulo $\kappa_\rho$. 
Next, choose $A_r\in \Conn(P_{f_r})$ with curvature $F_{A_r}=F_r$. Then, set
$$
A_c=\sum_rn^r(c)A_r.
\eqno(2.5.5)
$$
Similarly, choose $\Phi_r\in CS^2(M)$ with $c_{\Phi_r}=f_r$ and
$F_{\Phi_r}=F_r$ and $\Phi_\rho\in CS^2(M)$ with $c_{\Phi_\rho}=t_\rho$ 
and $F_{\Phi_\rho}=0$. As $\kappa_\rho t_\rho=0$, 
$\Phi_\rho{}^{\kappa_\rho}=\chi_a$ for some $a\in Z^1_{dR}(M)$, 
by the exact sequence (2.4.4). Redefining
$\Phi_\rho$ into $\Phi_\rho\chi_{a/\kappa_\rho}$, one can impose
$$
\Phi_\rho{}^{\kappa_\rho}=1.
\eqno(2.5.6)
$$
Then, set 
$$
\Phi_c=\prod_r\Phi_r{}^{n^r(c)}\cdot\prod_\rho\Phi_\rho{}^{k^\rho(c)}.
\eqno(2.5.7)
$$
Then, the maps $c\mapsto A_c$ and $c\mapsto\Phi_c$ are respectively
a connection and a compatible character assignment. 

Let a background connection assignment $c\mapsto A_c$ and a compatible 
background character assignment $c\mapsto \Phi_c$ be given. 
For $\Lambda\in Z^s_1(M)$ and $A\in\Conn(P_c)$, we define $\oint_\Lambda A$ 
by the procedure expounded in the previous subsection by taking 
$A_{P_c}=A_c$ and $\Phi_{P_c}=\Phi_c$, for $c\in H^2(M,\Bbb Z)$. 
In this way, (2.4.10)--(2.4.12) hold with $A_P$ and $\Phi_P$ replaced by 
$A_c$ and $\Phi_c$. It is convenient, though not necessary, to choose 
$A_c$, $\Phi_c$ of the form (2.5.5), (2.5.7).
In this way, the arbitrariness inherent in the definition
of $\oint_\Lambda A$, discussed at the end of the previous subsection, 
is reduced to that associated with the choice of $A_r$, $\Phi_r$, $\Phi_\rho$.

\vskip .3cm {\it 2.6  Example, the $4$--torus}

Since the formalism expounded above is rather abstract, we illustrate 
it with a simple example. We consider the case where $M$ is the $4$--torus
$T^4$. As coordinates of $T^4$, we use angles $\theta^i\in[0,2\pi[$, 
$1\leq i\leq 4$.

The $4$-torus $T^4$ has the nice property that torsion vanishes both in 
homology and in cohomology. Thus, we have the isomorphisms
$H^s_p(T^4)\cong H_{dR\Bbb Z}^p(T^4)\cong H^p(T^4,\Bbb Z)
\cong \Bbb Z^{C^4_p}$, where $C^4_p=b_p$ is a binomial coefficient.
A standard basis of $H^s_p(T^4)$ consists of the homology classes of the 
singular $p$--cycles $\Lambda_{a_1\cdots a_p}\in Z^s_p(T^4)$,
$1\leq a_1<\cdots <a_p \leq 4$, defined by
$$
\theta^i(t_1,\cdots, t_p)
=2\pi\sum_{s=1}^p\delta_{a_s}^i t_s,\quad 0\leq t_1,\cdots, t_p<1.
\eqno(2.6.1)
$$
A standard basis of $H_{dR\Bbb Z}^p(T^4)$ consists of the cohomology 
classes of the integer period $p$--forms $\omega^{a_1\cdots a_p}
\in Z_{dR\Bbb Z}^p(T^4)$, $1\leq a_1<\cdots <a_p \leq 4$,
defined by
$$
\omega^{a_1\cdots a_p}=\hbox{$1\over (2\pi)^p$}
d\theta^{a_1}\wedge \cdots\wedge d\theta^{a_p}.
\eqno(2.6.2)
$$
For a given $p$, the homology and cohomology basis are reciprocally dual. 

Since $H^2(T^4,\Bbb Z)\cong H_{dR\Bbb Z}^2(T^4)$, a principal $U(1)$ bundle 
on $T^4$ is determined up to equivalence by the de Rham cohomology class of 
the curvature of any connection. We consider the principal $U(1)$ bundle 
$P^{ab}\in\Princ(T^4)$ defined by the de Rham cohomology class of the 
$2$--form 
$$
F^{ab}=\omega^{ab}\in Z_{dR\Bbb Z}^2(T^4), 
\eqno(2.6.3)
$$
with $1\leq a<b\leq 4$. $P^{ab}$ is described concretely by the monodromy of a section of the associated line bundle around the $1$--cycles $\Lambda_c$
$$
T^{ab}{}_c=\exp(i\delta^a{}_c\theta^b-i\delta^b{}_c\theta^a)
\eqno(2.6.4)
$$
Any $P\in\Princ(T^4)$ is expressible as a product of 
$P^{ab}$'s and their inverses.
A connection $A^{ab}\in\Conn(P^{ab})$ with curvature $F^{ab}$ is
$$
A^{ab}=\hbox{$1\over 2(2\pi)^2$}\big(\theta^ad\theta^b
-\theta^bd\theta^a\big).
\eqno(2.6.5)
$$
$[F^{ab}]_{dR}\in H_{dR\Bbb Z}^2(T^4)$ determines 
unambiguously a class $c^{ab}\in H^2(T^4,\Bbb Z)$.
There is a unique Cheeger Simons character $\Phi^{ab}\in CS^2(T^4)$
such that $F_{\Phi^{ab}}=F^{ab}$, $c_{\Phi^{ab}}=c^{ab}$ and that
$$
\Phi^{ab}(\Lambda_c)=1, \quad 1\leq c\leq 4.
\eqno(2.6.6)
$$
Indeed, (2.6.6) selects unambiguously a unique character among those such that 
$F_{\Phi^{ab}}=F^{ab}$, $c_{\Phi^{ab}}=c^{ab}$ (cfr. the exact sequence 
(2.4.5)). By (2.4.1), (2.6.6) 
$$
\Phi^{ab}(\Lambda)=\exp\left(2\pi i\int_SF^{ab}\right),
\eqno(2.6.7)
$$
for $\Lambda=\sum_{a=1}^4 n_a\Lambda_a+b S\in H^s_1(T^4)$
with $n_a\in \Bbb Z$ and $S\in C^s_2(T^4)$ a $2$--chain.

A background connection assignment and a compatible background 
character assignment are given by
$$
A_c=\sum_{1\leq a<b\leq 4}n_{ab}(c)A^{ab},
\eqno(2.6.8)
$$
$$
\Phi_c=\prod_{1\leq a<b\leq 4}(\Phi^{ab})^{n_{ab}(c)},
\eqno(2.6.9)
$$
for $c=\sum_{1\leq a<b\leq 4}n_{ab}(c)c^{ab}\in H^2(T^4,\Bbb Z)$.

\vskip .3cm {\bf 3. The gauge partition function}

The physical motivation of the following construction has been given in the 
introduction.

To begin with, to properly define the kinetic term of the photon action
and to carry out the gauge fixing and quantization program, we endow $M$ 
with a fixed background Riemannian metric $g$.

\vskip .3cm {\it 3.1 The photon action}

For any $P\in \Princ(M)$ and any $A\in \Conn(P)$, the Wick rotated photon 
action $S(A,\tau)$ is given by 
\footnote{}{}\footnote{${}^1$}{The Wick rotated action $S$ is 
related to the Euclidean action $S_E$ as $S=iS_E$.} 
$$
S(A,\tau)=\pi\int_MF_A\wedge\hat\tau F_A.
\eqno(3.1.1)
$$
Here, $\tau$ varies in the open upper complex half plane $\Bbb H_+$, 
$$
\tau=\tau_1+i\tau_2,\quad \tau_1\in\Bbb R,\quad \tau_2\in\Bbb R_+
\eqno(3.1.2)
$$
and $\hat\tau$ is the operator
$$
\hat\tau=\tau_1+i*\tau_2.
\eqno(3.1.3)
$$
The action $S(A,\tau)$ takes the form (1.1) upon expressing $\tau$ as 
in (1.2) and rescaling $A$ into $(q/2\pi)A$. The integrality of the de Rham 
cohomology class of $F_A$ translates in the flux quantization condition (1.3)
after the rescaling.

The action $S(A,\tau)$ has the obvious symmetry
$$
A\rightarrow A+a
\eqno(3.1.4)
$$
where $a\in Z^1_{dR}(M)$. Unless $H^1(M,\Bbb R)=0$, this symmetry is 
larger than gauge symmetry, which corresponds to $a\in Z^1_{dR\Bbb Z}(M)$
(cfr. subsect. 2.2).

The field equations can be written as
$$
d\hat\tau F_A=0.
\eqno(3.1.5)
$$
They are equivalent to the vacuum Maxwell equations and the Bianchi identity
$$
d F_A=0, \quad d*F_A=0.
\eqno(3.1.6)
$$

\vskip .3cm {\it 3.2 The Wilson loop action}

The insertion of a Wilson loop along a cycle $\Lambda\in Z^s_1(M)$
is equivalent to add to the photon action a coupling of the gauge field 
$A$ to a one dimensional defect represented by $\Lambda$. For any 
$A\in\Conn(P)$, the interaction term of $A$ and $\Lambda$ is then
$$
W(A,\Lambda)=2\pi\oint_\Lambda A \qquad\hbox{mod $2\pi\Bbb Z$},
\eqno(3.2.1)
$$
where the right hand side is defined in the way expounded in subsect. 2.4.
The fact that $\Lambda$ is a $1$--cycle is equivalent to the conservation of 
the associated current. (See subsect. 3.5 below.)

As explained in subsect. 2.4, the definition of $\oint_\Lambda A$ involves 
choices and, thus, is not unique. It will be necessary to check at the end 
that the result of our calculations does not depend on the conventions 
used. 

\vskip .3cm {\it 3.3  The partition function}

The partition function with a Wilson loop insertion is given by
$$
Z(\Lambda,\tau)=\sum_{P\in \Princ(M)}
\int_{A\in\Conn(P)}
{DA\over {\rm vol}(\Gau(P))}
\exp\left(iS(A,\tau)+iW(A,\Lambda)\right)
\eqno(3.3.1)
$$
\ref{18,19,20,21}. The right hand side of this expression
is the formal mathematical statement 
of the physical quantization prescription consisting in a
summation over all topological classes of the gauge field and a 
functional integration of the quantum fluctuations of the gauge field 
about the vacuum gauge configuration of each class with the gauge group 
volume divided out.

To compute the above formal expression, we exploit heavily the results of 
subsect. 2.5. We first set $P=P_c$ with $c\in H^2(M,\Bbb Z)$ 
and transform the summation over $P$ into one over $c$.
Next, we choose a background connection assignment $c\mapsto A_c$
and write a generic $A\in \Conn(P_c)$ as
$$
A=A_c+v,
\eqno(3.3.2)
$$
where $v\in C^1_{dR}(M)$ is a fluctuation,
and transform the integration over $A$ into one over $v$.
To evaluate the Wilson loop action, we further pick 
a background character assignment $c\mapsto \Phi_c$ 
compatible with the connection assignment $c\mapsto A_c$.

It is possible and convenient to impose that the connections $A_c$
of the connection assignment satisfy the Maxwell equation
$$
d*F_c=0.
\eqno(3.3.3)
$$
To keep the arbitrariness involved in the various choices as controlled 
as possible,
we assume further that the background connection and character assignments 
$c\mapsto A_c$, $c\mapsto \Phi_c$ are of the form (2.5.5), (2.5.7),
respectively. 

Proceeding in this way, we find that the partition function factorizes in a 
classical background and a quantum fluctuation factor,
$$
Z(\Lambda,\tau)=Z_{\rm cl}(\Lambda,\tau)\cdot Z_{\rm qu}(\Lambda,\tau),
\eqno(3.3.4)
$$
where
$$
Z_{\rm cl}(\Lambda,\tau)=\sum_{c\in H^2(M,\Bbb Z)}
\exp\left(i\pi\int_MF_c\wedge\hat\tau F_c+2\pi i\oint_\Lambda A_c\right),
\eqno(3.3.5)
$$
$$
Z_{\rm qu}(\Lambda,\tau)
=\int_{v\in C^1_{\rm dR}(M)}
{\varrho Dv\over {\rm vol}(Z^1_{dR\Bbb Z}(M))}
\exp\left(-\pi\tau_2\int_Mdv\wedge*dv+2\pi i\oint_\Lambda v\right).
\eqno(3.3.6)
$$
$\varrho$ is a universal Jacobian relating the formal volumes 
${\rm vol}(\Gau(M))$ and ${\rm vol}(Z^1_{dR\Bbb Z}(M))$
(cfr. subsect. 2.2).

\vskip .3cm {\it 3.4 Evaluation of the classical partition function}

In order (3.3.3) to hold, the curvatures $F_r$ of the 
connections $A_r$ appearing in (2.5.5) all satisfy (3.3.3). Hence, 
the $F_r$ form a basis of the lattice $\Harm^2_{\Bbb Z}(M)$. 
The inverse intersection matrix $Q$ is defined by 
$$
Q_{rs}=\int_M F_r\wedge F_s.
\eqno(3.4.1)
$$
As well known, $Q$ is a unimodular symmetric integer $b_2\times b_2$ matrix
characterizing the topology of $M$ 
and $Q$ is even or odd according to whether $M$ is spin or not.
As $*\Harm^2(M)\subseteq\Harm^2(M)$ and $*^2=1$ on $\Harm^2(M)$, one has 
$$
*F_r=\sum_s H^s{}_r F_s,
\eqno(3.4.2)
$$
where $H$ is a non singular real $b_2\times b_2$ matrix such that $H^2=1$.
As $\int_M F\wedge *F$ is a norm on $\Harm^2(M)$,  
$QH$ is a positive definite symmetric $b_2\times b_2$ 
matrix.

From (2.5.5), one has immediately that
$$ 
F_c=\sum_rn^r(c)F_r.
\eqno(3.4.3)
$$

Recalling from subsect. 2.5 that
$\exp\left(2\pi i\oint_\Lambda A_c\right)=\Phi_c(\Lambda)$
and using (2.5.7), we find 
$$
\exp\left(2\pi i\oint_\Lambda A_c\right)=
\exp\left(2\pi i\sum_rn^r(c)\oint_\Lambda A_r\right)
\prod_\rho\Phi_\rho(\Lambda)^{k^\rho(c)}
\eqno(3.4.4)
$$

Using (3.3.5), (3.1.3), (3.4.1)--(3.4.4), we obtain
$$
Z_{\rm cl}(\Lambda,\tau)=\sum_{c\in H^2(M,\Bbb Z)}
\exp\left(i\pi n(c)^tQ(\tau_11+i\tau_2H)n(c)+2\pi in(c)^t\gamma(\Lambda)
\right)\prod_\rho\Phi_\rho(\Lambda)^{k^\rho(c)},
\eqno(3.4.5)
$$
where
$$
\gamma_r(\Lambda)=\oint_\Lambda A_r.
\eqno(3.4.6)
$$
From (2.5.4), by setting $n^r=n^r(c)$ and $k^\rho=k^\rho(c)$,
we can transform the summation over $c\in H^2(M,\Bbb Z)$ in a summation 
over $n^r\in \Bbb Z$ and $k^\rho=0,1,\ldots,\kappa_\rho-1$. 
Using (2.5.6), it is easy to see that
$$
\sum_{k^\rho=0,\ldots,\kappa_\rho-1}
\prod_\beta\Phi_\beta(\Lambda)^{k^\beta}=
\prod_\rho\kappa_\rho\varsigma(\Lambda),
\eqno(3.4.7)
$$
where the characteristic map $\varsigma$ is defined by
$$ 
\hbox{$\varsigma(\Lambda)=1$ ~~if $\Phi_\rho(\Lambda)=1$ for all $\rho$,
\quad $\varsigma(\Lambda)=0$~~ else.}
\eqno(3.4.8)
$$
Thus, 
$$
Z_{\rm cl}(\Lambda,\tau)=\sum_{n\in \Bbb Z^{b_2}}
\exp\left(i\pi n^tQ(\tau_11+i\tau_2H)n+2\pi in^t\gamma(\Lambda)\right)
\prod_\rho\kappa_\rho\varsigma(\Lambda),
\eqno(3.4.9)
$$
which is our final expression of the classical partition function.

The origin of the strange looking factor $\varsigma(\Lambda)$ 
is not difficult to interpret intuitively. 
Comparing (3.4.3), (3.4.4), we notice that, while the  
gauge curvature $F_c$ is not sensitive to the torsion part of $c$ 
(cfr. eq. (2.5.4)), the Abelian Wilson loop 
$\exp\left(2\pi i\oint_\Lambda A_c\right)$ is.
When we sum over all classes $c\in H^2(M,\Bbb Z)$ in (3.3.5), 
a finite subsum over all torsion classes $t\in\Tor^2(M,\Bbb Z)$
is involved. By (3.4.3), (3.4.4), the terms of the subsum
differ only by phases, which, on account of (2.5.6), are rational. 
The superposition of these phases leads to either constructive or 
destructive interference and yields the factor $\varsigma(\Lambda)$.
As explained in subsect. 2.4, the dependence of the Abelian Wilson loop  
$\exp\left(2\pi i\oint_\Lambda A_c\right)$ on the torsion part of $c$ can be 
traced to its dependence on the flat factors of the underlying 
principal bundle $P_c$. Thus, the factor $\varsigma(\Lambda)$
can ultimately be attribuited to an interference effect of the flat
topological classes in (3.3.1). 

\vskip .3cm {\it 3.5 Evaluation of the quantum partition function}

The computation of the quantum partition function proceeds through 
two basic steps \ref{26}. Firstly, one endows the relevant field spaces with 
suitable Hilbert structures in order to define the corresponding 
functional measures. Secondly, one determines the appropriate field 
kinetic operators required by the definition of the perturbative expansion. 
In our case, the problem is simplified by the fact that the field theory 
we are dealing with is free. There are however complications related to 
gauge invariance and the consequent need for gauge fixing. 

In our model, the relevant field spaces are $C^p_{dR}(M)$ with $p=0,~1$,
corresponding to the Faddeev--Popov ghost field and photon field. 
The Hilbert structure of $C^p_{dR}(M)$ is defined as usual by
$$
\langle u, v\rangle=\int_Mu\wedge *v,
\quad u,v\in C^p_{dR}(M).
\eqno(3.5.1)
$$
The relevant kinetic operators are the standard form
Laplacians $\Delta_p$ acting on $C^p_{dR}(M)$
$$
\Delta_p=(d^\dagger d+dd^\dagger)_p.
\eqno(3.5.2)
$$
which are order 2 elliptic non negative self adjoint operators. 

Since we are using a Hilbert space formalism, it is convenient
to express the argument of the exponential in (3.3.6) in terms 
of the Hilbert structure (3.5.1).
To this end, for a cycle $\Lambda\in Z^s_1(M)$, we define a
distribution $j_\Lambda$ on $C^1_{dR}(M)$ by 
$$
\langle j_\Lambda,\omega\rangle=\oint_\Lambda\omega,
\quad \omega\in C^1_{dR}(M).
\eqno(3.5.3)
$$ 
As a consequence of the relation $b\Lambda=0$, one has
$$
d^\dagger j_\Lambda=0.
\eqno(3.5.4)
$$
Intuitively, $j_\Lambda$ is the current associated to the 
$1$--cycle $\Lambda$ and (3.5.4) is the statement that $j_\Lambda$
is conserved.

As one is computing the partition function of a field theory on a generally
topologically non trivial manifold, particular care must be taken for a
proper treatment of the zero modes of the kinetic operators. 
The $p=0$ ghost zero modes form the $1$--dimensional vector space
of constant functions on $M$, $\Harm^0(M)$. 
As a basis of this, we choose the constant scalar $1$.
The $p=1$ photon zero modes form the $b_1$--dimensional vector space
of harmonic $1$--forms of $M$, $\Harm^1(M)$. 
As a basis of this, we choose a basis 
$\{\omega_m\}$, $m=1,\ldots,b_1$, of the lattice $\Harm^1_{\Bbb Z}(M)$
for convenience.

We fix the gauge by imposing the customary Lorentz fixing gauge condition. 
By using standard Faddeev--Popov type manipulations to perform the gauge
fixing, we find
$$
\eqalignno{
Z_{\rm qu}(\Lambda,\tau)=&
\left({\det G_1\over\vol M}\right)^{1\over 2}
{1\over (2\pi)^{b_1-1\over 2}}
\prod_n\delta_{\langle j_\Lambda,\omega_n\rangle,0}&\cr
\times &\left[\det{}'(\Delta_0){\det{}'(2\pi\tau_2\Delta_0)\over
\det{}'(2\pi\tau_2\Delta_1)}\right]^{1\over 2}
\exp\left(-\pi^2\langle j_\Lambda, 
(\pi\tau_2\Delta_1)^{-1}{}'j_\Lambda\rangle\right).&(3.5.5)\cr
}
$$
Here, $\det{}'(\Delta)$ and $\Delta^{-1}{}'$
denote the determinant and the inverse of the restriction of 
$\Delta$ to the orthogonal complement of its kernel, respectively, and 
$$
G_{1mn}=\langle\omega_m,\omega_n\rangle.
\eqno(3.5.6)
$$
We collect in the appendix the details of the derivation of (3.5.5).
Without going through all that, 
we can intuitively understand the origin of the various factors 
appearing in (3.5.5). $[\det{}'(2\pi\tau_2\Delta_1)]^{-{1\over 2}}$
is the photon determinant. Roughly speaking, the combination
$[\det{}'(\Delta_0)\det{}'(2\pi\tau_2\Delta_0)]^{1\over 2}$
is the ghost determinant, since the second determinant
equals the first up to a $\tau_2$ dependent constant.
The factor $\prod_n\delta_{\langle j_\Lambda,\omega_n\rangle,0}$
is yielded by the integration over the photon zero modes that satisfy the 
Lorentz gauge fixing condition with the volume of the residual gauge symmetry 
divided out.
The zero modes live in the torus $\Harm^1(M)/\Harm_{\Bbb Z}^1(M)$.
Only the integral $\exp\left(2\pi i\oint_\Lambda v\right)$ in (3.3.6)
depends on them. Integration of this phase on the zero modes torus 
produces the above combination of Kronecker delta functions. 
Finally, the exponential factor $\exp\left(-\pi^2\langle j_\Lambda, 
(\pi\tau_2\Delta_1)^{-1}{}'j_\Lambda\rangle\right)$ is the result of the 
Gaussian integration in (3.3.6) and represents the selfenergy of the current 
$j_\Lambda$. The remaining factors are just normalization constants.

In (3.5.5), both the determinants and the argument of the exponential suffer
ultraviolet divergences which have to be regularized and renormalized. 

We regularize the determinants using Schwinger's proper time method,
which now we briefly review \ref{26}.
Let $\Delta$ be an elliptic non negative self adjoint operator in some
Hilbert space of fields on a manifold $X$. Its proper time regularized
determinant is given by 
$$
\det{}'_\epsilon(\Delta)=\exp\left(-\int_\epsilon^\infty {dt\over t}
\big(\tr\exp(-t\Delta)-\dim\ker\Delta\big)\right),
\eqno(3.5.7)
$$
where $\epsilon>0$ is a small ultraviolet cut off of mass dimension exponent
$-2$. According to the heat kernel expansion
$$
\tr\exp(-t\Delta)\sim\sum_{k=0}^\infty
t^{k-\dim X\over\ord\Delta}\int_Xa_k(\Delta), \quad t\rightarrow 0+,
\eqno(3.5.8)
$$
where $a_k(\Delta)$ is a $\dim X$--form depending locally on the 
background geometry. Using (3.5.7), (3.5.8), it is easy to show that
$$
\eqalignno{
\det{}'_\epsilon(\Delta)=\epsilon^{-\dim\ker\Delta}\exp\Bigg\{&
-\sum_{l=1}^{\dim X}{\epsilon^{-l/\ord\Delta}\over l/\ord\Delta}
\int_Xa_{\dim X-l}(\Delta)&\cr
&+\ln\epsilon\int_Xa_{\dim X}(\Delta)+O(\epsilon)\Bigg\}
\det{}'_{\rm ms}(\Delta).&(3.5.9)\cr}
$$
Here, $\det{}'_{\rm ms}(\Delta)$ is the finite minimally subtracted 
renormalized determinant. 

We note that, for any $\kappa>0$, one has 
$$
\det{}'_\epsilon(\kappa\Delta)=\det{}'_{\kappa\epsilon}(\Delta),
\eqno(3.5.10)
$$
a simple property that will be useful in the calculations below. 

We replace the formally divergent determinants appearing in (3.5.5)
with their proper time regularized counterparts and use the expansion (3.5.9).
The expressions of the heat kernel forms $a_k(\Delta_p)$ are well known 
in the literature \ref{31}. In this way, one finds
$$
\eqalignno{
&\left[\det{}'_\epsilon(\Delta_0)
{\det{}'_\epsilon(2\pi\tau_2\Delta_0)\over
\det{}'_\epsilon(2\pi\tau_2\Delta_1)}\right]^{1\over 2}
=\epsilon^{{b_1\over 2}-1}
\exp\Bigg\{{1\over (8\pi)^2}\left({3\over (2\pi\tau_2)^2}-1\right)
{1\over \epsilon^2}\int_M d^4xg^{1\over 2}
&\cr
&
+{1\over (8\pi)^2}\left({1\over 2\pi\tau_2}+{1\over 3}\right)
{1\over \epsilon}\int_M d^4xg^{1\over 2}R
+{1\over (8\pi)^2}{1\over 90}\ln\epsilon\int_Md^4xg^{1\over 2}
\Big(25 R^2-88 R^{ij}R_{ij}
\vphantom{\Bigg\{}&\cr
&
+13 R^{ijkl}R_{ijkl}\Big)
+{1\over (8\pi)^2}{\ln(2\pi\tau_2)\over 60}
\int_Md^4xg^{1\over 2}
\Big(15 R^2-58 R^{ij}R_{ij}
+8 R^{ijkl}R_{ijkl}\Big)
+O(\epsilon)\Bigg\}\vphantom{\Bigg\{}&\cr
&\times
(2\pi\tau_2)^{b_1-1\over 2}{\det{}'_{\rm ms}(\Delta_0)\over
\det{}'_{\rm ms}(\Delta_1)^{1\over 2}}.&(3.5.11)\cr
}
$$
The prefactor $\epsilon^{{b_1\over 2}-1}$ can be absorbed into an 
appropriate $\epsilon$ dependent normalization of the zero mode part of 
the partition function measure. 
The local divergences appearing in the exponential can be removed by 
adding to the action $S(A,\tau)$ (cfr.  eq. (3.1.1)) 
local counterterms with suitable $\epsilon$ dependent 
coefficients. The general form of these counterterms, predicted also by 
standard power counting considerations, is
$$
\eqalignno{
\Delta S_\epsilon(\tau)
={i\over (8\pi)^2}\int_Md^4xg^{1\over 2}
\Big(&c_4(\epsilon,\tau)+c_2(\epsilon,\tau)R&\cr
+&\,\,c_0(\epsilon,\tau)R^2+c_0'(\epsilon,\tau)R^{ij}R_{ij}
+c_0''(\epsilon,\tau)R^{ijkl}R_{ijkl}\Big), \vphantom{\int}~~~~~
&(3.5.12)\cr
}
$$
where the suffix of the numerical coefficients denotes the exponent of 
their mass dimension.
If one adopts the minimal subtraction renormalization scheme, one obtains 
$$
\eqalignno{
\left[\det{}'(\Delta_0)
{\det{}'(2\pi\tau_2\Delta_0)\over
\det{}'(2\pi\tau_2\Delta_1)}\right]^{1\over 2}_{\rm ms}
=&\,\exp\Bigg\{{1\over (8\pi)^2}{\ln(2\pi\tau_2)\over 60}
\int_Md^4xg^{1\over 2}
\Big(15 R^2-58 R^{ij}R_{ij}\vphantom{\Bigg\{}&\cr
&~~~~+8 R^{ijkl}R_{ijkl}\Big)\Bigg\}
\times(2\pi\tau_2)^{b_1-1\over 2}{\det{}'_{\rm ms}(\Delta_0)\over
\det{}'_{\rm ms}(\Delta_1)^{1\over 2}}.~~~~~&(3.5.13)\cr
}
$$
As it turns out, 
the $\tau_2$ dependence of the resulting renormalized product of determinants 
has bad duality covariance properties due to the exponential factor. 
It is possible to remove the latter by adjusting the finite part of the 
local counterterms. This amounts to adopting another duality covariant
renormalization scheme for which
$$
\left[\det{}'(\Delta_0)
{\det{}'(2\pi\tau_2\Delta_0)\over
\det{}'(2\pi\tau_2\Delta_1)}\right]^{1\over 2}_{\rm dc}
=(2\pi\tau_2)^{b_1-1\over 2}{\det{}'_{\rm ms}(\Delta_0)\over
\det{}'_{\rm ms}(\Delta_1)^{1\over 2}}.
\eqno(3.5.14)
$$
It is Witten's choice \ref{18} and also ours.

Next, we regularize the Green function by using again Schwinger's proper time 
me\-thod, as described below \ref{26}. 
Let $\Delta$ be an elliptic non negative self adjoint 
operator in some Hilbert space of fields on a manifold $X$ as before. Its 
proper time regularized Green function is 
$$
\Delta^{-1}{}'{}_\epsilon
=\int_\epsilon^\infty dt\left(\exp(-t\Delta)-P(\ker\Delta)\right),
\eqno(3.5.15)
$$
where $P(\ker\Delta)$ is the orthogonal projector on $\ker\Delta$
and $\epsilon>0$ is a small ultraviolet cut off of mass dimension exponent
$-2$. Indeed, carrying out the integration explicitly, one has
$$
\Delta^{-1}{}'{}_\epsilon=\Delta^{-1}{}'\exp(-\epsilon\Delta).
\eqno(3.5.16)
$$

We note that, for any $\kappa>0$, one has 
$$
(\kappa\Delta)^{-1}{}'{}_\epsilon 
=\kappa^{-1}\Delta^{-1}{}'{}_{\kappa\epsilon},
\eqno(3.5.17)
$$
as is apparent also from (3.5.16).

The heat kernel $\exp(-t\Delta)(x,x')$, $x,~x'\in M$, is a bitensor with 
the small $t$ expansion
$$
\exp(-t\Delta)(x,x')\sim
{1\over (4\pi t)^{\dim X/2}}\exp\left(-{\sigma(x,x')\over 2t}\right)
\sum_{l=0}^\infty t^lf_l(x,x'),\quad t\rightarrow 0+.
\eqno(3.5.18)
$$
Here, $\sigma(x,x')$ is half the square geodesic distance of $x$, $x'$.
The $f_l(x,x')$ are certain bitensors of the same type as  
$\exp(-t\Delta)(x,x')$ \ref{31}.

We regularize the formal expression 
$\langle j_\Lambda,(\pi\tau_2\Delta_1)^{-1}{}'j_\Lambda\rangle$
appearing in (3.5.5) by replacing $(\pi\tau_2\Delta_1)^{-1}{}'$ with 
$(\pi\tau_2\Delta_1)^{-1}{}'{}_\epsilon$. The only thing one needs to know 
about the small $t$ expansion of the heat kernel $\exp(-t\Delta)_{ij'}(x,x')$
is that $f_{0ij'}(x,x')|_{x'=x}=g_{ij}(x)$ and 
$\partial_{k'}f_{0ij'}(x,x')|_{x'=x}=g_{kl}\Gamma^l_{ij}(x)$. 
In this way, one finds
$$
\langle j_\Lambda,(\pi\tau_2\Delta_1)^{-1}{}'{}_\epsilon j_\Lambda\rangle
={2\over (4\pi^2\tau_2)^{3\over 2}}{1\over \epsilon^{1\over 2}}
\int_0^1dt(\Lambda^*g_{tt})^{1\over 2}
+{1\over\pi\tau_2}\sigma(\Lambda)+O(\epsilon^{1\over 2}),
\eqno(3.5.19)
$$
where $\sigma(\Lambda)$ is a finite constant
depending on $\Lambda$. 
In the first term, the $1$--cycle $\Lambda$ is viewed as a parameterized
path $\Lambda:[0,1]\rightarrow M$ and the value of the integral is just
the length of the path as measured by the metric $g$.
The divergent part can be removed by adding to the interaction action
$W(A,\Lambda)$ (cfr. eq. (3.2.1)) a local counterterm of the form 
$$
\Delta W_\epsilon(\Lambda,\tau)
=i c_1(\epsilon,\tau)\int_0^1dt(\Lambda^*g_{tt})^{1\over 2}
\eqno(3.5.20)
$$
with a suitably adjusted $\epsilon$ dependent coefficient 
of mass dimension exponent $1$. 

One finds in this way
$$
Z_{\rm qu~ren}(\Lambda,\tau)=
\left({\det G_1\over\vol M}\right)^{1\over 2}
{\det{}'_{\rm ms}(\Delta_0)\over \det{}'_{\rm ms}(\Delta_1)^{1\over 2}}
\prod_n\delta_{\langle j_\Lambda,\omega_n\rangle,0}
\tau_2{}^{b_1-1\over 2}
\exp\left(-{\pi\sigma(\Lambda)\over \tau_2}\right),
\eqno(3.5.21)
$$
which is our final expression of the renormalized quantum partition function.
The factors appearing in (3.5.21) are easily interpreted. 
$\det{}'_{\rm ms}(\Delta_0)$, $\det{}'_{\rm ms}(\Delta_1)^{-{1\over 2}}$
are the renormalized ghost and photon determinants, respectively.
$\tau_2{}^{b_1-1\over 2}$ is the explicit $\tau_2$ dependence of the 
renormalized determinants. $\sigma(\Lambda)$ is the conventionally normalized 
renormalized selfenergy of the conserved current $j_\Lambda$ associated 
with $\Lambda$. The origin of the combination $\prod_n
\delta_{\langle j_\Lambda,\omega_n\rangle,0}$ was explained below (3.5.6).

\vskip .3cm {\it 3.6 Selection rules}

Let us examine the implications of the above calculation.
Consider a cycle $\Lambda\in Z^s_1(M)$.
From (3.4.8), (3.4.9), it follows that $Z_{\rm cl}(\Lambda,\tau)=0$ 
unless $\Phi_\rho(\Lambda)=1$ for all $\rho$, that is $\Lambda$ is 
contained in the kernel of all characters $\Phi\in CS^2(M)$
such that $c_\Phi\in\Tor^2(M,\Bbb Z)$. This is the classical selection rule. 
From (3.5.21), recalling that $\langle j_\Lambda,\omega_k\rangle
=\oint_\Lambda\omega_k$ by (3.5.3), it follows that 
$Z_{\rm qu}(\Lambda,\tau)=0$ unless 
$\oint_\Lambda\omega_k=0$ for all $k$, that is
$\Lambda$ is a torsion cycle, i. e. $[\Lambda]^s\in \Tor^s_1(M)$.  
This is the quantum selection rule.
From (3.3.4) and the above, we conclude  that $Z(\Lambda,\tau)=0$
identically unless $\Lambda\in Z^s_1(M)$ satisfies
$$
[\Lambda]^s\in \Tor^s_1(M),
\eqno(3.6.1)
$$
$$
\Phi(\Lambda)=1, \quad \hbox{for all $\Phi\in CS^2(M)$ with $c_\Phi=0$}.
\eqno(3.6.2)
$$

\vskip .3cm {\it 3.7 Flat characters and the Morgan--Sullivan 
torsion invariant}

Let $\Phi\in CS^2(M)$ be a flat character, i. e.  such that $F_\Phi=0$. Then, 
$c_\Phi\in \Tor^2(M,\Bbb Z)\cong\Tor^2_{s\Bbb Z}(M)$. Therefore, 
there exist a minimal integer $\nu_\Phi\in\Bbb N$, an integer cocycle 
$\rho\in Z^2_{s\Bbb Z}(M)$ and an integer cochain $s\in C^1_{s\Bbb Z}(M)$ 
such that $c_\Phi=[\rho]_{s\Bbb Z}$ and $\nu_\Phi\rho=ds$. 
On the other hand, as explained in subsect. 2.4, there is a real cochain 
$f\in C^1_{s\Bbb R}(M)$ such that $\Phi
=\exp\left(2\pi i f\big|Z^s_1(M)\right)$, $df\in Z^2_{s\Bbb Z}(M)$ and 
$c_\Phi=[d f]_{s\Bbb Z}$. We thus have, $df=\rho+dt$ for some 
integer cochain $t\in C^1_{s\Bbb Z}(M)$.

Let $\Lambda\in Z^s_1(M)$ such that $[\Lambda]^s\in \Tor^s_1(M)$.
Then, there are a minimal $\nu_\Lambda\in\Bbb N$ and 
$S\in C^s_2(M)$ such that $\nu_\Lambda \Lambda =bS$.

Using the above relations, one easily shows that 
$\nu_\Lambda f(\Lambda)=\rho(S)+\nu_\Lambda t(\Lambda)$
and $\nu_\Phi\rho(S)=\nu_\Lambda s(\Lambda)$.
Thus
$$
f(\Lambda)=\rho(S)/\nu_\Lambda=s(\Lambda)/\nu_\Phi
\quad \hbox{mod $\Bbb Z$}.
\eqno(3.7.1)
$$

Now, using (3.7.1), it is easy to check that $f(\Lambda)$
depends only on the cohomology class $c_\Phi$ of $\rho$
and the homology class $[\Lambda]^s$ of $\Lambda$
mod $\Bbb Z$. Hence, the object defined by 
$$
\langle [\Lambda]^s, c_\Phi\rangle=f(\Lambda)\quad \hbox{mod $\Bbb Z$}
\eqno(3.7.2)
$$
is a topological invariant. It is called Morgan--Sullivan torsion invariant
pairing \ref{32,33}. It is $\Bbb Z$ linear in both arguments and non singular.

From the above, we conclude that, for a character $\Phi\in CS^2(M)$
such that $F_\Phi=0$,
$$
\Phi(\Lambda)=\exp\left(2\pi i\langle [\Lambda]^s, c_\Phi\rangle\right),
\eqno(3.7.3)
$$
for all $\Lambda\in Z^s_1(M)$ such that $[\Lambda]^s\in\Tor^s_1(M)$.

\vskip .3cm {\it 3.8 The final form of the selection rules}

Using the results of the previous subsection, we can restate the selection 
rules (3.6.1), (3.6.2) as follows: 
$$
[\Lambda]^s\in \Tor^s_1(M),
\eqno(3.8.1)
$$
$$
\langle [\Lambda]^s, c\rangle=0 \quad \hbox{mod $\Bbb Z$},
\quad c\in \Tor^2_{s\Bbb Z}(M).
\eqno(3.8.2)
$$
As the Morgan--Sullivan pairing is non singular, these are equivalent to
$$
\Lambda\in B^s_1(M).
\eqno(3.8.2)
$$

Thus, {\it the partition function $Z(\Lambda, \tau)$ vanishes unless 
$\Lambda$ is a $1$--boundary}. This is the final form of the selection rules
of the Abelian Wilson loops. Note that {\it they originate from 
a non trivial combination of the classical and quantum selection rules.}

Remarkably, {\it in spite of the ambiguity inherent in the definition of the 
integral  $\oint_\Lambda A$, the partition function 
$Z(\Lambda,\tau)$ is unambiguously defined.} Indeed, 
as explained in subsect. 2.4, the indetermination of 
$\oint_\Lambda A$ is of the form $\oint_\Lambda a$ mod $\Bbb Z$ with 
$a\in Z^1_{dR}(M)$ and this object vanishes when $\Lambda$ is a boundary.
When, conversely, $\Lambda$ is not a boundary, $Z(\Lambda,\tau)$ 
vanishes identically, regardless the way the ambiguity of $\oint_\Lambda A$
is fixed.

This selection rule found is rather surprising when compared to the result for 
Abelian Chern Simons theory \ref{34}, where non trivial Abelian Wilson 
loops are found for non trivial knots. This calls for an explanation.
As a gauge theory on a topologically non trivial manifold $M$, 
Chern Simons theory is rather trivial, since the underlying principal bundle 
is trivial. For non trivial bundles, the 
Chern Simons Lagrangian would not be globally defined on $M$ in general
and thus could not be integrated to yield an action. 
Further, it is implicitly assumed that there are no photon zero modes. 
This restricts the manifold $M$ to be such that $H^1(M,\Bbb R)=0$. 
Thus, unlike for Maxwell theory, the quantization of Chern Simons theory
involves no sum over the topological classes of the gauge field, 
since only the trivial class is involved.
For this reason, the basic interference mechanism involving flat bundles
which is partly responsible for the selection rule of Abelian Wilson loops 
of Maxwell theory is not working in Abelian Chern Simons theory.
Further, all $1$--cycles $\Lambda$ one is dealing with are torsion 
from the start. 
Finally, in the Abelian Chern Simons model the relevant invariants 
of a knot are given in terms of the selfenergy of the current asociated 
to the knot, which is of a topological nature. In Maxwell 
theory, the self energy of a $1$--cycle is obviously not topological.

\vskip .3cm {\it 3.9  Example, the $4$--torus}

We illustrate the above results with an example. We consider again
the case where $M$ is $4$--torus $T^4$, which was already discussed 
in subsect. 2.6.

It is not difficult to compute the $\tau$ dependent factor of the 
partition function $Z(\Lambda,\tau)$. $Z(\Lambda,\tau)$ is given 
by (3.3.4) with $Z_{\rm cl}(\Lambda,\tau)$, $Z_{\rm qu}(\Lambda,\tau)$ 
given respectively by (3.4.9), (3.5.21) (after renormalization).  
Since $\Tor^2(T^4,\Bbb Z)=0$, the factor 
$\prod_\rho\kappa_\rho\varsigma(\Lambda)$ appearing in (3.4.9)
is identically $1$. The Betti numbers $b_1$, $b_2$ of the $4$--torus $T^4$
are $4$, $6$, respectively. It follows that, for a $1$--boundary 
$\Lambda\in B^s_1(T^4)$,  
$$
Z(\Lambda,\tau)=Z_0\tau_2{}^{3\over 2}
\exp\left(-{\pi\sigma(\Lambda)\over \tau_2}\right)
\Psi(\gamma(\Lambda),\tau),
\eqno(3.9.1)
$$
where $Z_0$ is a constant independent from $\Lambda$, $\tau$,
$\gamma(\Lambda)$ is defined in (3.4.6) and $\Psi(\gamma,\tau)$ 
is a certain function  of $\gamma\in\Bbb C^6$, $\tau\in\Bbb H_+$,
given by (3.4.9) with $\gamma(\Lambda)$ 
replaced by $\gamma$ and $\prod_\rho\kappa_\rho\varsigma(\Lambda)$
set to $1$. 

It is not difficult to compute $\Psi(\gamma,\tau)$ when 
$T^4$ is endowed with the standard flat metric
$$
g=\delta_{ij}d\theta^i\otimes d\theta^j.
\eqno(3.9.2)
$$
The $2$--forms $\omega^{ab}$, $1\leq a<b\leq 4$, 
defined in (2.6.2), belong to $\Harm_{\Bbb Z}^2(T^4)$ and form a basis
of this latter. A simple calculations shows that 
$Q^{ab,cd}=\int_{T^4}\omega^{ab}\wedge\omega^{cd}=\epsilon^{abcd}$ and 
$QH^{ab,cd}=\int_{T^4}\omega^{ab}\wedge*\omega^{cd}=\delta^{ac}\delta^{bd}
-\delta^{ad}\delta^{bc}$. If we use the index $r=1,2,3,4,5,6$
for the pairs $(ab)=(12),(34),(13),(24),(14),(23)$, $Q$ and $QH$ 
are representable as the $6\times 6$ matrices 
$$
Q=\sigma_1\oplus-\sigma_1\oplus\sigma_1, \quad
QH=1_2\oplus1_2\oplus1_2,
\eqno(3.9.3)
$$
where $1_2$ is the $2\times 2$ unit matrix and $\sigma_1$ is a Pauli matrix.
Using (3.9.3), it is straightforward to show that
$$
\Psi(\gamma,\tau)
=\psi(\gamma^{(1)},\tau)\psi(\gamma^{(2)},-\bar\tau)
\psi(\gamma^{(3)},\tau),
\quad \gamma=\gamma^{(1)}\oplus\gamma^{(2)}\oplus\gamma^{(3)}
\eqno(3.9.4)
$$
where $\gamma^{(h)}\in \Bbb C^2$ and, for $\tau\in\Bbb H_+$, $g\in\Bbb C^2$, 
$$
\psi(g,\tau)
=\vartheta_2(g_1+g_2|2\tau)\overline{\vartheta_2(\bar g_1-\bar g_2|2\tau)}
+\vartheta_3(g_1+g_2|2\tau)\overline{\vartheta_3(\bar g_1-\bar g_2|2\tau)},
\eqno(3.9.5)
$$
$\vartheta_2$, $\vartheta_3$ being standard Jacobi theta functions.

\vskip .3cm {\bf 4. Analysis of Abelian duality}

We now come to the analysis of the duality covariance properties of 
the partition function with Wilson loop insertion $Z(\Lambda,\tau)$, which 
is the main subject of he paper. 

\vskip .3cm {\it 4.1 Study of the $\tau$ dependence and duality}

\def\thfc#1#2#3#4{\vartheta_{#4}\left[{#1\atop #2}\right]\left(#3\right)}
We next study the $\tau$ dependence of the partition function
$Z(\Lambda,\tau)$. This resides essentially in a $\vartheta$ function
of the appropriate characteristics. It is therefore necessary to review first 
some of the basics of the theory of $\vartheta$ functions.
See for instance \ref{35} for background.

We recall that the standard $\vartheta$ function with characteristics
is defined by
$$
{\thfc x y K b}=\sum_{n\in\Bbb Z^b+x}\exp\left(i\pi n^t Kn+2\pi in^ty\right),
\eqno(4.1.1)
$$
where $b\in \Bbb N$, $x,~y\in\Bbb R^b$ and $K\in\Bbb C(b)$ 
such that $K=K^t$ and $\imag K>0$.
The main properties of ${\thfc x y K b}$ used below are the following.
Using the Poisson resummation formula, one can show that 
the $\vartheta$ function satisfies the relation
$$
{\thfc x y K b}=\det(-iK)^{-{1\over 2}}\exp\left(2\pi ix^ty\right)
{\thfc y {-x} {-K^{-1}} b},
\eqno(4.1.2)
$$
where the branch of the square root used is that for which 
$u^{1\over 2}>0$ for $u>0$. 
If $L\in\Bbb R(b)$ induces an automorphism of the lattice $\Bbb Z^b$, 
one has
$$
{\thfc x y K b}={\thfc {L^{-1}x} {L^ty} {L^tKL} b}.
\eqno(4.1.3)
$$
An element  $Z\in \Bbb Z(b)$ with $Z=Z^t$ is said even if $n^tZn\in 2\Bbb Z$ 
for any $n\in \Bbb Z^b$ and odd else. We set $\nu_Z=1$ if $Z$ is even and
$\nu_Z=2$ if $Z$ is odd. Then, one has 
$$
{\thfc x y K b}=\exp\left(\nu_Z\pi ix^tZx\right)
{\thfc x {y-\nu_ZZx} {K+\nu_ZZ} b}
\eqno(4.1.4)
$$

From (3.3.4), (3.4.9), (3.5.21), the $\tau$ dependent factor of 
the partition function $Z(\Lambda,\tau)$ can be written as
$$
{\cal Z}(\Lambda,\tau)=
\exp\left(-{\pi\sigma(\Lambda)\over\tau_2}\right){\cal F}(\Lambda,\tau),
\eqno(4.1.5)
$$
where
$$
{\cal F}(\Lambda,\tau)=\tau_2{}^{b_1-1\over 2}
{\thfc 0 {\gamma(\Lambda)} {K(\tau)} {b_2}}.
\eqno(4.1.6)
$$
Here, $\tau=\tau_1+i\tau_2$ varies in the open upper complex half plane
$\Bbb H_+$. On account of the selection rules derived in subsect. 3.8, 
we can assume tha $\Lambda\in B^s_1(M)$ is a boundary. 
$K(\tau)$ is given by 
$$
K(\tau)=Q(\tau_1+i\tau_2H),
\eqno(4.1.7)
$$
where $Q$ and $H$ are defined by (3.4.1), (3.4.2), respectively.
Since $Q, H\in\Bbb R(b_2)$, $Q=Q^t$, $QH=(QH)^t$ and $QH>0$,
$K(\tau)\in\Bbb C(b_2)$, $K(\tau)=K(\tau)^t$ and $\imag K(\tau)>0$,
as required. The vector $\gamma(\Lambda)\in\Bbb R^{b_2}$ 
is given by (3.4.6). $\gamma(\Lambda)$ is defined modulo $\Bbb Z^{b_2}$.
Since $\Lambda$ is a boundary and the curvatures $F_r$ of the connections 
$A_r$ satisfy the Maxwell equations (3.3.3), $\gamma(\Lambda)$ does not 
depend on the choice of the $A_r$ modulo $\Bbb Z^{b_2}$.
For convenience, we have extracted the exponential factor 
$\exp(-\pi\sigma(\Lambda)/\tau_2)$, whose $\tau$ dependence
is anyway quite simple.

The analysis of duality reduces essentially to the study of the covariance 
properties of the function ${\cal Z}(\Lambda,\tau)$ under a suitable subgroup 
of the modular group \ref{18,19}, whose main properties we now briefly 
review \ref{36}. 

The modular group $\bar\Gamma[1]$ consists of all 
transformations of the open upper complex half plane $\Bbb H_+$ of the form
$$
u(\tau)={a\tau+b\over c\tau+ d}, \quad \hbox{with $a, b, c, d\in \Bbb Z$,
$ad-bc=1$}. 
\eqno(4.1.8)
$$
As is well known, $\bar\Gamma[1]$ is generated by two elements 
$s$, $t$ defined by
$$
s(\tau)=-1/\tau, \quad t(\tau)=\tau+1.
\eqno(4.1.9)
$$
These satisfy the relations
$$
s^2=\id,\quad (st)^3=\id.
\eqno(4.1.10)
$$
The modular group $\bar\Gamma[1]$ is isomorphic to the group
$\PSL(2,\Bbb Z)\cong\SL(2,\Bbb Z)/\{-1,1\}$, the isomorphism being
defined by
$$
A(u)=\pm\left(\matrix{a&b&\cr
c&d&\cr}\!\!\!\!\!\!\right),
\eqno(4.1.11)
$$
with $u\in\bar\Gamma[1]$ given by (4.1.18). In particular, 
$$
A(s)=\pm\left(\matrix{0&-1&\cr
1&0&\cr}\!\!\!\!\!\!\right),\qquad
A(t)=\pm\left(\matrix{1&1&\cr
0&1&\cr}\!\!\!\!\!\!\right).
\eqno(4.1.12)
$$

To efficiently study the duality covariance of ${\cal F}(\Lambda,\tau)$, 
it is necessary to introduce a class of functions of $\tau\in\Bbb H_+$ 
defined as follows. Recall that $Q\in\Bbb Z(b_2)$ and $Q=Q^t$ and, 
so, $Q$ can be even or odd (according to whether $M$ is spin or not). 
For $k,~l\in \Bbb Z$ with $kl\in \nu_Q\Bbb Z$, we set
$$
{\cal F}_{(k,l)}(\Lambda,\tau)=\tau_2{}^{b_1-1\over 2}
\exp\left(-i\pi kl\gamma(\Lambda)^tQ^{-1}\gamma(\Lambda)\right)
{\thfc {kQ^{-1}\gamma(\Lambda)} {l\gamma(\Lambda)} {K(\tau)} {b_2}}.
\eqno(4.1.13)
$$
It is readily checked that this expression is defined unambiguously 
in spite of the $\Bbb Z^{b_2}$ indeterminacy of $\gamma(\Lambda)$.
Our function ${\cal F}(\Lambda,\tau)$ is actually a member of this
function class, since indeed 
$$
{\cal F}(\Lambda,\tau)={\cal F}_{(0,1)}(\Lambda,\tau).
\eqno(4.1.14)
$$

A simple analysis shows that 
$$
{\cal F}_{(k,l)}(\Lambda,\tau)=
e^{{i\pi\over 4}\eta}
\tau^{-{\chi+\eta\over 4}}\bar\tau^{-{\chi-\eta\over 4}}
{\cal F}_{(l,-k)}(\Lambda,-1/\tau).
\eqno(4.1.15)
$$
Here, $\chi$ and $\eta$ are respectively the Euler and signature invariant
of $M$ and are given by
$$
\chi=2(1-b_1)+b_2,
\eqno(4.1.16)
$$
$$
\eta=b_2^+-b_2^-.
\eqno(4.1.17)
$$
To prove (4.1.15), one uses (4.1.2), (4.1.3) with $L=Q$,
and the relations $b_2=b_2^++b_2^-$ and 
$$
\det\left(-iK(\tau)\right)^{1\over 2}=
e^{-{i\pi\over 4}\eta}
\tau^{b_2^+/2}\bar\tau^{b_2^-/2},
\eqno(4.1.18)
$$
$$
-K(\tau)^{-1}=Q^{-1}K(-1/\tau)Q^{-1}.
\eqno(4.1.19)
$$

Using (4.1.4), one shows similarly that
$$
{\cal F}_{(k,l)}(\Lambda,\tau)={\cal F}_{(k,l-\nu_Q k)}(\Lambda,\tau+\nu_Q).
\eqno(4.1.20)
$$

Let $G_{\nu_Q}$ be the subgroup of $\bar\Gamma[1]$ generated by $s$ and 
$t^{\nu_Q}$. Specifically, $G_1=\bar\Gamma[1]$ and $G_2=\bar\Gamma_\theta$, 
the so called Hecke subgroup of $\bar\Gamma[1]$. In \ref{18,19}, it was shown
that $G_{\nu_Q}$ is the duality group, the subgroup  of $\bar\Gamma[1]$ under 
which the partition function without insertions behaves as a modular form of 
weights $\chi+\eta\over 4$, $\chi-\eta\over 4$. Now, (4.1.14) and (4.1.20)
can be written as
$$
\eqalignno{
{\cal F}_{(k,l)}(\Lambda,\tau)=&\,e^{{i\pi\over 4}\eta}
\tau^{-{\chi+\eta\over 4}}\bar\tau^{-{\chi-\eta\over 4}}
{\cal F}_{(k,l)A(s)^{-1}}(\Lambda,s(\tau))&\cr
=&\,{\cal F}_{(k,l)A(t^{\nu_Q})^{-1}}(\Lambda,t^{\nu_Q}(\tau)).
&(4.1.21)\cr}
$$
Since ${\cal F}_{(k,l)}(\Lambda,\tau)={\cal F}_{(-k,-l)}(\Lambda,\tau)$, 
as is easy to show from (4.1.13) using (4.1.1), the above expressions are 
unambiguously defined in spite of the sign indeterminacy of 
$A(s)$ and $A(t^{\nu_Q})$. (4.1.21) states that {\it 
${\cal F}_{(k,l)}(\Lambda,\tau)$ is a generalized modular form 
of $G_{\nu_Q}$ of weights $\chi+\eta\over 4$, $\chi-\eta\over 4$.
In this sense, $G_{\nu_Q}$ continues to be the 
duality group also for the partition function with Wilson loop insertions.}
%(More about this in the next subsection.)

We denote by $E_{\nu_Q}(\Lambda)$ the subspace of $\Fun(\Bbb H_+)$
spanned by the functions ${\cal F}_{(k,l)}(\Lambda,\tau)$.
We note that, when $\gamma(\Lambda)$ satisfies certain restrictions, the 
functions ${\cal F}_{(k,l)}(\Lambda,\tau)$ are not all independent. For 
instance,
if $\gamma(\Lambda)=0$ mod $\Bbb Z^{b_2}$, ${\cal F}_{(k,l)}(\Lambda,\tau)$
is actually independent from $k,~l$. So, $E_{\nu_Q}(\Lambda)$ may in some
instance be finite dimensional. To see how this can come about in greater 
detail, suppose that $\gamma(\Lambda)\in\Bbb Q^{b_2}$. 
Then, there is a minimal $p\in\Bbb N$ such that $p\gamma(\Lambda)\in 
\Bbb Z^{b_2}$. Let $k,~l\in \Bbb Z$ such that $kl\in\nu_Q\Bbb Z$. 
Let further $m,~n\in \Bbb Z$ such that $(kn+lm+mnp)p\in\nu_Q\Bbb Z$. Then, 
$(k+mp)(l+np)\in\nu_Q\Bbb Z$ and, as is easy to show from (4.1.13), one has
$$
{\cal F}_{(k+mp,l+np)}(\Lambda,\tau)=
\exp\left(2\pi i(nk-ml-mnp)w(\Lambda)/\nu_Qp\right)
{\cal F}_{(k,l)}(\Lambda,\tau),
\eqno(4.1.22)
$$
where $w(\Lambda)\in \Bbb Z$ is given by
$$
w(\Lambda)=\hbox{$1\over 2$}\nu_Qp^2\gamma(\Lambda)^tQ^{-1}\gamma(\Lambda).
\eqno(4.1.23)
$$
The phase factor is a $\nu_Qp$--th root of unity independent from $\tau$.
Therefore, when $\gamma(\Lambda)$ satisfies the above condition, 
$E_{\nu_Q}(\Lambda)$ is finite dimensional. A standard basis 
of $E_{\nu_Q}(\Lambda)$ consists of the ${\cal F}_{(k,l)}(\Lambda,\tau)$ such 
that $0\leq k,l\leq p-1$. The dimension of $E_{\nu_Q}(\Lambda)$ is therefore 
$$
n_p=p^2-[p/2]^2(\nu_Q-1).
\eqno(4.1.24)
$$

Denote by ${\cal F}_A(\Lambda,\tau)$ the standard basis of 
$E_{\nu_Q}(\Lambda)$. Combining (4.1.15), (4.1.20) and (4.1.22), it is simple 
to show that there are invertible $n_p\times n_p$ complex matrices 
$S_{AB}(\Lambda)$ and $T^{\nu_Q}{}_{AB}(\Lambda)$ such that 
$$
{\cal F}_A(\Lambda,\tau)=
e^{{i\pi\over 4}\eta}
\tau^{-{\chi+\eta\over 4}}\bar\tau^{-{\chi-\eta\over 4}}
\sum_BS_{AB}(\Lambda){\cal F}_B(\Lambda,-1/\tau),
\eqno(4.1.25)
$$
$$
{\cal F}_A(\Lambda,\tau)=\sum_B
T^{\nu_Q}{}_{AB}(\Lambda){\cal F}_B(\Lambda,\tau+\nu_Q).
\eqno(4.1.26)
$$
This means that ${\cal F}_A(\Lambda,\tau)$ is the $A$-th component of a vector
modular form ${\cal F}(\Lambda,\tau)$
of $G_{\nu_Q}$ of weights $\chi+\eta\over 4$, $\chi-\eta\over 4$.

The matrices $S_{AB}(\Lambda)$ and $T^{\nu_Q}{}_{AB}(\Lambda)$ have the 
property that only one matrix element in each row and column is non zero. 
For instance, if $p=2$ and $\nu_Q=1$, one has $n_p=4$, 
$A=(0,0),~(0,1),~(1,0),~(1,1)$ and 
$$
S(\Lambda)=\left(\,\,\matrix{
1&  0 &  0 &  0 &\cr
0&  0 &  1 &  0 &\cr
0&  1 &  0 &  0 &\cr
0&  0 &  0 &  \varepsilon_\Lambda&\cr
}\!\!\!\!\!\right),\quad
T(\Lambda)=\left(\,\,\matrix{
1&  0 &  0 &  0 &\cr
0&  1 &  0 &  0 &\cr
0&  0 &  0 &  \varepsilon_\Lambda  &\cr
0&  0 &  1 &  0&\cr
}\!\!\!\!\!\right),\qquad\varepsilon_\Lambda=\exp(-i\pi w(\Lambda)).
\eqno(4.1.27)
$$
For $p=2$, $\nu_Q=2$, one has $n_p=3$, 
$A=(0,0),~(0,1),~(1,0)$ and 
$$
S(\Lambda)=\left(\,\,\matrix{
1& 0 & 0 &\cr
0& 0 & 1 &\cr
0& 1 & 0 &\cr
}\!\!\!\!\!\right),\quad
T^2(\Lambda)=\left(\,\,\matrix{
1& 0 & 0 &\cr
0& 1 & 0 &\cr
0& 0 & \varepsilon_\Lambda &\cr
}\!\!\!\!\!\right),\qquad\varepsilon_\Lambda=\exp(-i\pi w(\Lambda)/2).
\eqno(4.1.28)
$$

\vskip .3cm {\it 4.2 Duality and Twisted sectors}

The question arises whether the formal considerations expounded in the
previous subsection have a physical interpretation. Here, we propose one.

To anticipate, {\it to each boundary $\Lambda\in B^s_1(M)$, there is 
associated a family ${\scri T}_\Lambda$ of twisted sectors of the quantum 
field theory. ${\scri T}_\Lambda$ is characterized by a point of the 
cohomology torus $H_{dR}^2(M)/H_{dR\Bbb Z}^2(M)$ and is parameterized by
a pair of integers $k,~l\in\Bbb Z$ such that $kl\in\nu_Q\Bbb Z$
and satisfying further restrictions when $\gamma(\Lambda)\in\Bbb Q^{b_2}$,
as explained earlier. In turn, each sector is a collection of topological 
vacua in one--to--one correspondence with $\Princ(M)$, as usual.
The $\tau$ dependent factor of the partition function with a Wilson loop 
insertion associated to $\Lambda$ of the sector $k,l$ is
$$
{\cal Z}_{(k,l)}(\Lambda,\tau)=
\exp\left(-{\pi\sigma(\Lambda)\over\tau_2}\right){\cal F}_{(k,l)}(\Lambda,\tau)
\eqno(4.2.1)
$$
(cfr. eq. (4.1.5)).} In the rest of the subsection, we shall try to justify 
the claims made.

For $\Lambda\in B^s_1(M)$, we define first
$$
B_\Lambda=\sum_{rs}Q^{-1rs}\left(\oint_\Lambda A_r\right)A_s.
\eqno(4.2.2)
$$
$$
G_\Lambda=dB_\Lambda=\sum_{rs}Q^{-1rs}\left(\oint_\Lambda A_r\right)F_s.
\eqno(4.2.3)
$$
As is easy to see from (3.4.1),
$$
\oint_\Lambda B_\Lambda=\int_M G_\Lambda\wedge G_\Lambda.
\eqno(4.2.4)
$$

Next, for $k,~l\in\Bbb Z$ with $kl\in\nu_Q\Bbb Z$, we define the action 
$$
\eqalignno{
S_{(k,l)}(A,\Lambda,\tau)=&\,\,\pi\int_M(F_A+kG_\Lambda)\wedge
\hat\tau(F_A+kG_\Lambda)
+2\pi l\oint_\Lambda (A+kB_\Lambda)&\cr
&\,-\pi kl\int_M G_\Lambda\wedge G_\Lambda, &(4.2.5)\cr}
$$
where $A\in\Conn(P)$ with $P\in\Princ(M)$
(cfr. eqs. (3.1.1)--(3.1.3) and (3.2.1)). We shall consider now the 
quantum field theory defined by $S_{(k,l)}(A,\Lambda,\tau)$. But first a few 
remarks are in order.

Since $\oint_\Lambda A_r\not\in\Bbb Z$ is defined up to an arbitrary 
integer $m_r$, $B_\Lambda$ is defined up to a shift of the form 
$B_m=\sum_{rs}Q^{-1rs}m_rA_s$. Correspondingly, $G_\Lambda$ is defined up to 
a shift of the form $G_m=\sum_{rs}Q^{-1rs}m_rF_s$. 
Note that $B_m$ is a connection of a $U(1)$ principal bundle $Q_m$ such that 
$n^r(c_{Q_m})=\sum_{rs}Q^{-1rs}m_s$ (cfr. eqs. (2.5.4), (2.5.5)) and that 
$G_m$ is its curvature.

If we make the replacements $B_\Lambda\rightarrow B_\Lambda +B_m$ and 
$G_\Lambda\rightarrow G_\Lambda +G_m$, one has
$$
S_{(k,l)}(A,\Lambda,\tau)\rightarrow S_{(k,l)}(A+kB_m,\Lambda,\tau)
+\pi kl\int_M G_m\wedge G_m.
\eqno(4.2.6)
$$
Note that $A+kB_m\in\Conn(PQ_m{}^k)$. Further, $kl\int_M G_m\wedge G_m\in 
2\Bbb Z$. 

Next, we come to  the quantum field theory defined by the action 
$S_{(k,l)}(A,\Lambda,\tau)$. Its partition function 
is computed summing over all topological vacua of $\Princ(M)$ and factoring 
the classical and quantum fluctuation contributions, as usual.
As is easy to see, the ambiguity (4.2.6) is absorbed by exponentiation  
and topological vacua summation. 

A calculation completely analogous to that expounded in sect. 3
for the partition function $Z(\Lambda,\tau)$ shows that
the $\tau$ dependent factor of the partition 
function is precisely ${\cal Z}_{(k,l)}(\Lambda,\tau)$, eq (4.2.1).

The class of $G_\Lambda$ in $Z^2_{dR}(M)$ modulo $Z^2_{dR\Bbb Z}(M)$ is the 
point of $H_{dR}^2(M)/H_{dR\Bbb Z}^2(M)$ characterizing ${\scri T}_\Lambda$
mentioned at the beginning of the subsection.

The conclusion of the analysis is that, {\it to preserve Abelian duality in 
the presence of Wilson loops, it is necessary to assume the
existence of twisted sectors.}

\vskip .3cm {\bf Appendix}

In this appendix, we provide briefly the details of the derivation of the 
formal expression (3.5.5) of the quantum partition function
$Z_{\rm qu}(\Lambda,\tau)$. The starting expression of 
$Z_{\rm qu}(\Lambda,\tau)$, given in (3.3.6), is a formal 
functional integral which requires a careful treatment.

We normalize conventionally the functional measure $D\varphi$
on a Hilbert space $\scri F$ of fields $\varphi$ so that
$$
\int_{\varphi\in {\scri F}}D\varphi
\exp\left(-\hbox{$1\over 2$}\langle \varphi, 
\varphi\rangle\right)=1.
\eqno(A.1)
$$

In our case, the relevant field Hilbert spaces are certain subspaces of 
$C^p_{dR}(M)$ with $p=0,~1$ equipped with the Hilbert space structure 
defined by (3.5.2). The corresponding functional measures are characterized
by (A.1).

The invariant measure on the gauge group $\Gau(M)$ is defined by the 
translation of that on its Lie algebra $\Lie\Gau(M)$ once the normalization 
of the exponential map is chosen. Recall that $\Lie\Gau(M)\cong C^0_{dR}(M)$. 
We fix the normalization by writing $h\in\Gau(M)$ near $1$ as $h
=\exp(2\pi i f)$ with $f\in C^0_{dR}(M)$ and choose $Df$ as the measure on 
$\Lie\Gau(M)$.

Let us go back to (3.3.6). We fix the gauge by imposing a generalized 
Lorentz condition
$$
d_1{}^\dagger v=a, \quad v\in C^1_{dR}(M), 
\eqno(A.2)
$$
where $a\in\ran d_1{}^\dagger$ t. We then employ a slight variant of the 
Faddeev--Popov trick. 

We define a functional $B(v,a)$ of the fields $v\in C^1_{dR}(M)$, 
$a\in \ran d_1{}^\dagger$ through the identity
$$
1=B(v,a)\int_{x\in\ran d_0}Dx\,
\delta_{\ran d_1{}^\dagger}(d_1{}^\dagger(v+x)-a).
\eqno(A.3)
$$
It is easy to show that
$$
B(v-x,a)=B(v,a), \quad x\in\ran d_0.
\eqno(A.4)
$$
Further, when $v$ satisfies the gauge fixing condition (A.2),
$$
B(v,a)=B_0,
\eqno(A.5)
$$
where $B_0$ is a constant.
We now insert these relations in the functional integral (3.3.6) and, 
after some straightforward manipulations, we obtain 
$$
\eqalignno{ 
Z_{\rm qu}(\Lambda,\tau)
={\varrho B_0
\over {\rm vol}(\Harm_{\Bbb Z}^1(M))}\int_{v\in C^1_{\rm dR}(M)}&\,Dv 
\,\delta_{\ran d_1{}^\dagger}(d_1{}^\dagger v-a)&\cr
&\times\exp\left(-\langle v,(\pi\tau_2d^\dagger d)_1v\rangle
+2\pi i\langle j_\Lambda, v\rangle\right),
\vphantom{\int_{v\in C^1_{\rm dR}(M)}}~~~~~&(A.6)\cr}
$$
where $j_\Lambda$ is defined in (3.5.3).
Here, we have used the identity $\ran d_0=B^1_{dR}(M)$ and 
the formal relation
$$
{\rm vol}(Z^1_{dR\Bbb Z}(M))/{\rm vol}(B^1_{dR}(M))=
{\rm vol}(\Harm_{\Bbb Z}^1(M)).
\eqno(A.7)
$$

Next, we define a function $\Gamma(\xi)$ of the parameter $\xi>0$
by the formal identity
$$
1=\Gamma(\xi)\int_{a\in \ran d_1{}^\dagger}Da\exp(-\xi\langle a, a\rangle).
\eqno(A.8)
$$
Introducing the above relation in the functional integral (A.6), we eliminate 
the $\delta$ function, obtaining
$$
Z_{\rm qu}(\Lambda,\tau)
={\varrho B_0\Gamma(\xi)
\over {\rm vol}(\Harm_{\Bbb Z}^1(M))}\int_{v\in C^1_{\rm dR}(M)}Dv 
\exp\Big(-\langle v,(\pi\tau_2d^\dagger d+\xi dd^\dagger)_1v\rangle
+2\pi i\langle j_\Lambda, v\rangle\Big).
\eqno(A.9)
$$

We compute first the Jacobian $\varrho$. Recalling the facts about the 
structure of the gauge group $\Gau(M)$ expounded in subsect. 2.2, we 
find the formal relation
$$
\varrho=\vol(Z^1_{dR\Bbb Z}(M))/\vol(\Gau(M))
=\vol(B^1_{dR}(M))/\vol(\Gau_c(M)).
\eqno(A.10)
$$
The tangent map of the isomorphism $\alpha:\Gau_c(M)/\Gau_0(M)
\rightarrow B^1_{dR}(M)$ at the identity is just $d_0|_{\ker d_0{}^\perp}$.
From here. we have 
$\varrho=\det{}'\left((d^\dagger d)_0\right)^{1\over 2}/\vol(G(M))$.
One easily computes
$\vol(G(M))=\left(\vol M/2\pi\right)^{1\over 2}$.
Thus,
$$
\varrho=\left[{2\pi\det{}'\left((d^\dagger d)_0\right)
\over \vol M}\right]^{1\over 2}.
\eqno(A.11)
$$

The constant $B_0$ is easily computed from (A.3), taking (A.2) into account
and writing $x=df$ with $f\in \ker d_0{}^\perp$.
The result is 
$$
B_0=\det{}'\left((d^\dagger d)_0\right)^{1\over 2}.
\eqno(A.12)
$$

Similarly, $\Gamma(\xi)$ is easily computed from (A.8), writing 
$a=d_1{}^\dagger x$ with $x\in\ker d_1{}^{\dagger\perp}$:
$$
\Gamma(\xi)=\left[{\det{}'\left(2\xi (dd{}^\dagger)_1\right) \over 
\det{}'\left((dd^\dagger)_1\right)}\right]^{1\over 2}.
\eqno(A.13)
$$

The functional integrand (A.9) is invariant under the shifts $v\rightarrow 
v+\hat v_0$, where $\hat v_0\in \Harm_{\Bbb Z}^1(M)$, as is easy to see.
Thus, we can factorize the functional integration as follows
$$
{1\over {\rm vol}(\Harm_{\Bbb Z}^1(M))}\int_{v\in C^1_{\rm dR}(M)}Dv
=\int_{v_0\in\Harm^1(M)/\Harm_{\Bbb Z}^1(M)}Dv_0
\int_{v'\in\Harm^1(M)^\perp}Dv'.
\eqno(A.14)
$$
Proceeding in this way, we carry out the Gaussian integration 
straightforwardly and obtain
$$
\eqalignno{ 
&\int_{v\in C^1_{\rm dR}(M)}Dv 
\exp\Big(-\langle v,(\pi\tau_2d^\dagger d+\xi dd^\dagger)_1v\rangle
+2\pi i\langle j_\Lambda, v\rangle\Big)=
\left({\det G_1\over (2\pi)^{b_1}}\right)^{1\over 2}
\prod_k\delta_{\langle j_\Lambda,\omega_k\rangle,0}&\cr
&~~~~~~~~~~~~~
\times\det{}'\left((2\pi\tau_2d^\dagger d
+2\xi dd^\dagger)_1\right)^{-{1\over 2}}
\exp\left(-\pi^2\langle j_\Lambda, 
(\pi\tau_2d^\dagger d+\xi dd^\dagger)_1{}^{-1}{}'j_\Lambda\rangle\right),
&(A.15)\cr}
$$
where $G_1$ is the matrix given by (3.5.6).

Next, we substitute (A.11), (A.12), (A.13) and (A.15) into (A.9).
The resulting expression can be simplified noting that
the operators $(d^\dagger d)_0$, $(dd^\dagger)_1$
have the same non zero spectrum counting also multiplicity
and, thus, equal determinants and that
$$
\det{}'\left((pd^\dagger d+q dd^\dagger)_1\right)
=\det{}'\left(p(d^\dagger d)_1\right)
\det{}'\left(q(dd^\dagger)_1\right),
\eqno(A.16)
$$
with $p,~q>0$. Proceeding in this way, the $\xi$ gauge independence of 
$Z_{\rm qu}(\Lambda,\tau)$ becomes manifest and one straightforwardly 
obtains (3.5.5).

\vskip.6cm
\par\noindent
{\bf Acknowledgments.} We are greatly indebted to R. Stora for useful 
discussions. This paper is dedicated to the memory of my grandmother 
Ornella Scaramagli, whose loving and serene eyes still stare at me 
in my heart.

\par\noindent
{\bf Note added.} After this paper was published in Communication 
in Mathematical Physics, we became aware of refs.\ref{37,38}, where
the string $S$ duality, studied in \ref{14},  was conjectured for the 
first time.

\vskip.6cm
\centerline{\bf REFERENCES}
\vskip.6cm

\item{\ref{1}}
P.~A.~M.~Dirac,
``Quantized Singularities in the Electromagnetic Field'',
Proc.\ Roy.\ Soc.\ {\bf A133} (1931) 60.

\item{\ref{2}}
T.~ T.~Wu and C.~N.~ Yang,
``Concept of non Integrable Phase Factors and Global Formulation of Gauge 
Fields''.
Phys.\ Rev.\ {\bf D12} (1975) 3845.

\item{\ref{3}}
%\cite{Schwinger:nj}\bibitem{Schwinger:nj}
J.~S.~Schwinger,
``Magnetic Charge and Quantum Field Theory'',
Phys.\ Rev.\  {\bf 144} (1966) 1087.
%%CITATION = PHRVA,144,1087;%%

\item{\ref{4}}
%\cite{Zwanziger:by}\bibitem{Zwanziger:by}
D.~Zwanziger,
``Exactly Soluble Nonrelativistic Model of Particles with Both Electric and 
Magnetic Charges'',
Phys.\ Rev.\  {\bf 176} (1968) 1480.
%%CITATION = PHRVA,176,1480;%%

\item{\ref{5}}
%\cite{Polyakov:ek}\bibitem{Polyakov:ek}
A.~M.~Polyakov,
``Particle Spectrum In Quantum Field Theory'',
JETP Lett.\  {\bf 20} (1974) 194
[Pisma Zh.\ Eksp.\ Teor.\ Fiz.\  {\bf 20} (1974) 430].
%%CITATION = JTPLA,20,194;%%

\item{\ref{6}}
%\cite{'tHooft:1974qc}\bibitem{'tHooft:1974qc}
G.~'t Hooft,
``Magnetic Monopoles in Unified Gauge Theories'',
Nucl.\ Phys.\  {\bf B79} (1974) 276.
%%CITATION = NUPHA,B79,276;%%

\item{\ref{7}}
%\cite{Julia:ff}\bibitem{Julia:ff}
B.~Julia and A.~Zee,
``Poles with Both Magnetic and Electric Charges in Nonabelian Gauge Theory'',
Phys.\ Rev.\  {\bf D11} (1975) 2227.
%%CITATION = PHRVA,D11,2227;%%

\item{\ref{8}}
%\cite{Prasad:kr}\bibitem{Prasad:kr}
M.~K.~Prasad and C.~M.~Sommerfield,
``An Exact Classical Solution for the 'T Hooft Monopole and the 
Julia-Zee Dyon'',
Phys.\ Rev.\ Lett.\  {\bf 35} (1975) 760.
%%CITATION = PRLTA,35,760;%%

\item{\ref{9}}
E.~B.~Bogomolnyi,
``The Stability of Classical Solutions'',
Sov.\ J.\ Nucl.\ Phys.\ {\bf 24} (1976) 449.

\item{\ref{10}}
%\cite{Witten:mh}\bibitem{Witten:mh}
E.~Witten and D.~I.~Olive,
``Supersymmetry Algebras that Include Topological Char\-ges'',
Phys.\ Lett.\  {\bf B78} (1978) 97.
%%CITATION = PHLTA,B78,97;%%

\item{\ref{11}}
%\cite{Seiberg:1994aj}\bibitem{Seiberg:1994aj}
N.~Seiberg and E.~Witten,
``Monopoles, Duality and Chiral Symmetry Breaking in N=2 Supersymmetric QCD'',
Nucl.\ Phys.\  {\bf B431} (1994) 484,
arXiv:hep-th/9408099.
%%CITATION = HEP-TH 9408099;%%

\item{\ref{12}}
%\cite{Cremmer:1979up}\bibitem{Cremmer:1979up}
E.~Cremmer and B.~Julia,
``The SO(8) Supergravity'',
Nucl.\ Phys.\ {\bf B159} (1979) 141.
%%CITATION = NUPHA,B159,141;%%

\item{\ref{13}}
%\cite{Gaillard:1981rj}\bibitem{Gaillard:1981rj}
M.~K.~Gaillard and B.~Zumino,
``Duality Rotations for Interacting Fields'',
Nucl.\ Phys.\  {\bf B193} (1981) 221.
%%CITATION = NUPHA,B193,221;%%

\item{\ref{14}}
%\cite{Font:1990gx}\bibitem{Font:1990gx}
A.~Font, L.~E.~Ibanez, D.~Lust and F.~Quevedo,
``Strong--Weak Coupling Duality and Nonperturbative Effects in String Theory'',
Phys.\ Lett.\  {\bf B249} (1990) 35.
%%CITATION = PHLTA,B249,35;%%

\item{\ref{15}}
%\cite{Hull:1994ys}\bibitem{Hull:1994ys}
C.~M.~Hull and P.~K.~Townsend,
%``Unity of superstring dualities,''
Nucl.\ Phys.\ B {\bf 438} (1995) 109,
arXiv:hep-th/9410167.
%%CITATION = HEP-TH 9410167;%%

\item{\ref{16}}
%\cite{Witten:1995ex}\bibitem{Witten:1995ex}
E.~Witten,
``String Theory Dynamics in Various Dimensions'',
Nucl.\ Phys.\ {\bf B443} (1995) 85,
arXiv:hep-th/9503124.
%%CITATION = HEP-TH 9503124;%%

\item{\ref{17}}
%\cite{Strominger:1995cz}\bibitem{Strominger:1995cz}
A.~Strominger,
``Massless Black Holes and Conifolds in String Theory'',
Nucl.\ Phys.\  {\bf B451} (1995) 96,
arXiv:hep-th/9504090.
%%CITATION = HEP-TH 9504090;%%

\item{\ref{18}}
%\cite{Witten:1995gf}\bibitem{Witten:1995gf}
E.~Witten, 
``On S Duality in Abelian Gauge Theory'',
arXiv:hep-th/9505186.
%%CITATION = HEP-TH 9505186;%%

\item{\ref{19}}
%\cite{Verlinde:1995mz}\bibitem{Verlinde:1995mz}
E.~Verlinde,
``Global Aspects of Electricmagnetic Duality'',
Nucl.\ Phys.\ {\bf B455} (1995) 211,
arXiv:hep-th/9506011.
%%CITATION = HEP-TH 9506011;%%

\item{\ref{20}}
%\cite{Olive:2000yy}\bibitem{Olive:2000yy}
D.~I.~Olive and M.~Alvarez,
``Spin and Abelian Electromagnetic Duality on Four-Mani\-folds'',
Commun.\ Math.\ Phys.\  {\bf 217} (2001) 331,
arXiv:hep-th/0003155.
%%CITATION = HEP-TH 0003155;%%

\item{\ref{21}}
%\cite{Alvarez:1999uq}\bibitem{Alvarez:1999uq}
M.~Alvarez and D.~I.~Olive,
``The Dirac Quantization Condition for Fluxes on Four-Manifolds'',
Commun.\ Math.\ Phys.\  {\bf 210} (2000) 13
arXiv:hep-th/9906093.
%%CITATION = HEP-TH 9906093;%%

\item{\ref{22}}
%\cite{Olive:pt}\bibitem{Olive:pt}
D.~I.~Olive,
``Exact Electromagnetic Duality'',
prepared for NATO Advanced Study Institute on Strings, 
Branes and Dualities, Cargese, France, 26 May - 14 Jun 1997.

\item{\ref{23}}
%\cite{Alvarez-Gaume:sj}\bibitem{Alvarez-Gaume:sj}
L.~Alvarez-Gaume and F.~Zamora,
``Duality in Quantum Field Theory (and String Theory)'',
prepared for 37th Internationale Universitatswochen fuer Kernphysik und 
Teilchenphysik: Broken Symmetries (37 IUKT), Schladming, Austria, 
28 Feb - 7 Mar 1998, arXiv:hep-th/9709180.

\item{\ref{24}}
%\cite{Olive:2001xy}\bibitem{Olive:2001xy}
D.~I.~Olive,
``Spin and Electromagnetic Duality: An outline'',
arXiv:hep-th/0104062.
%%CITATION = HEP-TH 0104062;%%

\item{\ref{25}}
R.~Bott and L.~Tu,
``Differential Forms in Algebraic Topology'',
Springer Verlag, New York, 1982.

\item{\ref{26}}
%\cite{Schwarz:cb}\bibitem{Schwarz:cb}
A.~S.~Schwarz,
``Quantum Field Theory And Topology'',
Springer Verlag, Berlin, 1993.

\item{\ref{27}}
J.~-L.~Brylinski,
``Loop Spaces, Characteristic Classes and Geometric Quantization'',
Birkh\"auser, 1993.

\item{\ref{28}}
J.~L.~ Koszul, 
``Travaux de S. S. Chern et J. Simons sur les Classes Caract\'eristiques''
Seminaire Bourbaki 26\`eme ann\'ee {\bf 440} (1973/74) 69.

\item{\ref{29}}
J.~Cheeger,
``Multiplication of Differential Characters'',
Convegno Geometrico INDAM, Roma maggio 1972, in 
Symposia Mathematica {\bf XI} Academic Press (1973) 441.

\item{\ref{30}}
J.~Cheeger and J.~Simons,
``Differential Characters and Geometric Invariants'',
Stony Brook preprint (1973) reprinted in 
Lecture Notes in Math. {\bf 1167} Sprin\-ger Verlag (1985) 50.

\item{\ref{31}}
%\cite{Gilkey:mj}\bibitem{Gilkey:mj}
P.~B.~Gilkey,
``Invariance Theory, the Heat Equation and the Atiyah-Singer Index Theorem'',
Publish or Perish, Wilmington, 1984.					

\item{\ref{32}}
J.~W.~Morgan and D.~P.~ Sullivan
``The Transversality Characteristic Class and Linking Cycles in Surgery 
Theory'',
Ann.\ Math.\ {\bf 99} (1974) 461.
		
\item{\ref{33}}
%\cite{Freed:zx}\bibitem{Freed:zx}
D.~S.~Freed,
``Determinants, Torsion and Strings'',
Commun.\ Math.\ Phys.\  {\bf 107} (1986) 483.
%%CITATION = CMPHA,107,483;%%

\item{\ref{34}}
%\cite{Witten:1988hf}
E.~Witten,
``Quantum Field Theory and the Jones Polynomial,''
Commun.\ Math.\ Phys.\  {\bf 121} (1989) 351.
%%CITATION = CMPHA,121,351;%%

\item{\ref{35}}
J.~Igusa,
``Theta Functions'',
Springer Verlag, Berlin 1972.

\item{\ref{36}}
T.~Miyake
``Modular Forms'',
Springer Verlag, Berlin 1989.

\item{\ref{37}}
S.~J.~Rey,
``The Confining Phase of Superstrings and Axionic Strings'',
Phys.\ Rev.\ D {\bf 43} (1991) 526.
%%CITATION = PHRVA,D43,526;%%

\item{\ref{38}}
S.~J.~Rey,
``Axionic String Instantons And Their Low-Energy Implications'',
preprint UCSB-TH-89/49
{\it Invited talk given at Workshop on Superstrings and Particle Theory, Tuscaloosa, 
Alabama, Nov 8-11, 1989}.

\bye